%% file: sc_main-noformat.tex
\documentclass[conference]{IEEEtran}
\usepackage{cite}
\usepackage{amsmath,amssymb,amsfonts}
\usepackage{algorithmic}
\usepackage{graphicx}
\usepackage{textcomp}
\usepackage{xcolor}
\def\BibTeX{{\rm B\kern-.05em{\sc i\kern-.025em b}\kern-.08em T\kern-.1667em\lower.7ex\hbox{E}\kern-.125emX}}
\usepackage[utf8]{inputenc}
\usepackage{url}
\usepackage{color, soul}
\usepackage{multirow}
\usepackage{multicol}
\allowdisplaybreaks[4]
\begin{document} 
\title{DaSGD: Squeezing SGD Parallelization Performance in Distributed Training Using Delayed Averaging}

\author{
	\IEEEauthorblockN{
\begin{tabular} {c c c c}
	\multicolumn{4}{c} {\begin{tabular}{c c c}
			Qinggang~Zhou\textsuperscript{1*} & Yawen~Zhang\textsuperscript{2*} & Pengcheng Li\textsuperscript{1} \\
			\small{qinggangz@gmail.com} & \small{zhywenwen@pku.edu.cn} & \small{pengcheng.li@alibaba-inc.com} \\ \end{tabular}} \\
	\multicolumn{4}{c}{ }\\
	Xiaoyong Liu\textsuperscript{1} & Jun Yang\textsuperscript{1}& Runsheng Wang\textsuperscript{2} & Ru Huang\textsuperscript{2} \\
	\small{xiaoyong.liu@alibaba-inc.com} & \small{muzhuo.yj@alibaba-inc.com}& \small{r.wang@pku.edu.cn}&\small{ ruhuang@pku.edu.cn} \\
	\multicolumn{4}{c}{ }\\
\end{tabular} 
}
	\IEEEauthorblockA{
	\textsuperscript{1}\textit{Alibaba Group, Sunnyvale, USA} \\
	\textsuperscript{2}\textit{Peking University, Beijing, P.R. China} \\
	\textsuperscript{*} Both authors contributed equally to this work  }
}

\maketitle

\begin{abstract}
	The state-of-the-art deep learning algorithms rely on distributed training systems to tackle the increasing sizes of models and training data sets.
    Minibatch stochastic gradient descent (SGD) algorithm requires workers to halt forward/back propagations, to wait for gradients aggregated from all workers, and to receive weight updates before the next batch of tasks.
    This synchronous execution model exposes the overheads of gradient/weight communication among the large number of workers a distributed training system.
     We propose a new SGD algorithm, DaSGD (Local SGD with Delayed Averaging), which parallelizes SGD and forward/back propagations to hide $100\%$ of the communication overhead.
	By adjusting the gradient update scheme, this algorithm uses hardware resources more efficiently and reduces the reliance on the low-latency and high-throughput inter-connects. 
	The theoretical analysis and the experimental results show its convergence rate $ O (1 / \sqrt {K} )$, the same as SGD. 
	The performance evaluation demonstrates it enables a linear performance scale-up with the cluster size.

\end{abstract}


\input{introduction.tex}
\input{background.tex}
\input{theoretical.tex}

\input{experimental.tex}

\input{system.tex}

\input{strategy.tex}

\input{conclusion.tex}
\newpage
\bibliography{IEEEabrv,sample-base}
\bibliographystyle{IEEEtran}
\input{appendix.tex}

\end{document}

%% file: introduction.tex
\section{Introduction}
Training deep learning models using data parallelism on a large-scale distributed cluster has become an effective method for deep learning model training. 
The enormous training data set allows a huge batch of training tasks on different data samples running in parallel. As a result, the training task can be scaled out to a massive number of servers (workers). 
The pinnacle of this method reduces the training time of the benchmark ResNet-50 from days to a couple of minutes.  \cite{goyal2017accurate, you2017imagenet, akiba2017extremely, you2017scaling,ying2018image}
However, during the Mini-batch stochastic gradient descent (SGD) at the end of a batch, these workers have to halt, wait for the computed gradients aggregated from all of the workers and receive a weight update before starting the next batch. The wait time tends to worsen when the number of workers increases. Additionally, as the workloads are spread over a larger cluser, the computation time are greatly shorten and the communication overheads take a larger portion of the overall cost. 

System designers address this concern by improving inter-chip connects with higher throughput and lower latency and refining network topology \cite{li2019evaluating}, such as NVIDIA DGX-1 \cite{NVIDIADGX1} and NVIDIA DGX-2 \cite{NVSwitch}. Additional care has been given to reduce the intermediate steps that would increase communication latency. These methods effectively reduce the wait time during Mini-batch SGD on a large-scale distributed system \cite{Gaudi}. 

A modern data center design prefers selecting cost-efficient hardware blocks and choosing a balanced configuration for the typical workloads \cite{barroso2018datacenter}. Under these workloads, various hardware resources would be utilized in a balanced fashion. A distributed training system works in the opposite manner. During the forward propagation and back propagation phases, the computing resources are throttled at the peak throughputs while the system inter-connects and switches are completely idle. During the SGD phase, the forward propagation and back propagation tasks of the next batch are blocked from starting. So, the computing resources are mostly idle while the system inter-connects and switches are throttled at the peak throughputs. Improving system efficency over the communication cost may be archieved from an orthogonal direction of improving system inter-connects. That is, the workloads may be restructured or re-designed for a balanced utilization of the system hardware resources.

Inspired by the modern system design practices, we propose a new SGD method called DaSGD, enabling SGD running parallelly with forward/back propagation.
It replaces a Mini-batch SGD with Local SGD iterations to serialize forward/back propagations of different samples and to allow inter-worker weight averages may merge with local weights between Local SGD iterations.
Model averaging may be scheduled to be delayed for a limited number of Local SGD iterations, which hides communication time on a large distributed cluster.
Based on the network throughput and the data amount that training a model needs to tranfer, this algorithm may adjust the delay amount. This algorithm makes better use of distributed training systems and reduces the reliance on low latency and high peak throughput communication hardware.
The theoretical analysis clarifies its convergence rate is $ O (1 / \sqrt {K} )$, the same as the traditional SGD. 
The auxiliary parameters are added to realize quantitative control, and their proper ranges and design guidelines are also provided in exprimental results.
Finally, the system evaluation results show that this algorithm enables performance scale-up linearly with cluster size and is not restricted by communication.


The main contributions of this proposal are the followings.
\begin{itemize}
\item We present a new gradient aggregation algorithm for a large-scale deep learning training system, called DaSGD. This algorithm enables a more balanced and better utilized distributed training system.
\item We provide the theoretical analysis of the algorithm’s convergence rate. It shows the proposed algorithm converges at $ O (1 / \sqrt {K} )$, the same as regular SGD.
\item Our experiments show within the reasonable parameter ranges, this algorithm allows the training converges at the same rate of SGD. The experiments also explore the proper ranges of these parameters.
\item A performance evaluation of real-life systems reflects the impacts from many specific design issues in the system hardware and software stacks. These include but not limited to the communication scheduling in software framework, the reduction algorithm, GPU interconnect topology and interfaces, server interconnect topology and interfaces. 
They introduce unneccessary complexity and are out of the scope of our discussion. 
Instead, we abstract an analytical model using the key performance parameters based on the system configuration and the training setup. We show the system evaluation demonstrates the method produces a linear scale of efficiency with the cluster size.
\item A framework and further discussions are provided that guides how to use the method based on the system configuration and the training setup for best results. 
\end{itemize}
 
The context of this paper is structured as follows: Session~\ref{section:background} describes the background of distributed training and the related work about SGD, Session~\ref{section:theoretical} presents the design framework, the theoretical analysis of convergence rate and the discussion about the guidance scheme of DaSGD, Session~\ref{section:experimental} shows the exprimental results, Session~\ref{section:system} provides the system evaluation results, Session~\ref{section:strategy} discusses the training system design strategies, Session~\ref{section:conclusion} gives a conclusion.

%% file: background.tex
\section{Background and Related Work}\label{section:background}

\subsection{Stochastic Gradient Descent}
Stochastic Gradient Descent (SGD) is the backbone of numerous deep learning algorithms~\cite{Ghadimi2013Stochastic}. Supervised deep learning demands massive training datasets and super dense neural network architectures. Training a deep learning model needs many epochs for training to converge. A variant of classic SGD, synchronous mini-batch SGD~\cite{bottou2010large}, has become the mainstream, supported by prevalent machine learning frameworks, such as Tensorflow~\cite{Abadi2016Tensor}, Pytorch~\cite{Paszke2019pytorch}, MxNet~\cite{chen2015mxnet}. 

It computes gradients from a batch of training samples, as shown in Eq.~\ref{equ:updaterule_minibatchsgd3}.
\begin{equation}\label{equ:updaterule_minibatchsgd3} 
	x_{k+1}  = x_{k} - \frac{\eta}{B} \sum\limits_{j=1}^{B}\nabla F(x_{k},s_{k}^{(j)} )  
\end{equation} 
\noindent
where $x\in\mathbb{R}^{d}$ is the weight of model, $\eta$ is the learning rate, $B$ is the batch size, $\mathcal{S}$ is the training dataset, $s_{k}^{(j)}\subset\mathcal{S}$ is a random sample, $\nabla F(x_{k},s_{k}^{(j)} )$ is the stochastic gradient of the loss function of the sample $s_{k}^{(j)}$.


From a system perspective, a distributed training system may compute a batch of gradients on all  workers. At the end of a batch, a reduction operation is performed on the gradients on a worker first and a worker sends out only a copy of local averaged gradients. Further reductions are performed on gradients from different workers until a final copy of the averaging gradients is obtained. The above equation may be rewritten as
$x_{k+1}  = x_{k} - \frac{\eta}{M} \sum\limits_{j=1}^{B}g(x_{k},s_{k}^{(j)} )  $.
where  $M$ is the number of workers,  $g(x_{k},s_{k}^{(j)} )$ is the stochastic gradient that worker $j$ aggregates locally for that batch.$g(x_{k},s_{k}^{(j)} )=\frac{M}{B}\sum_{i=1}^{\frac{B}{M}}\nabla(x_{k},s_{k}^{(i)})$.

\subsection{Distributed Training Process based on SGD}

\begin{figure}[t]
	\centering
	\includegraphics[width=0.8\linewidth]{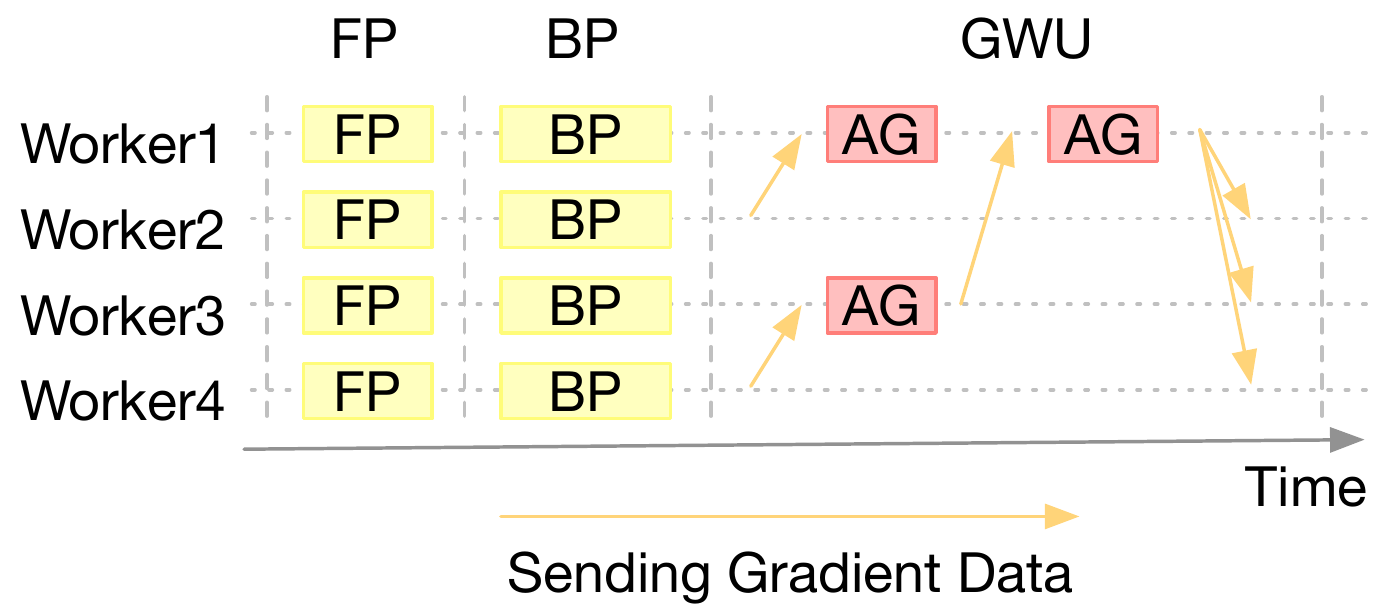}
	\caption{The SGD timeline of multi-worker distributed training based on data parallelism. FP, BP, GWU, and AG represent forward propagation, backward propagation, global weight update, and gradient averaging.}
	\label{fig:timeline}
\end{figure}

The training of a neural network is an iterative process, and the \emph{weight}s of a neural network layer need to be computed frequently. Each computation does the following phases sequentially: forward propagation, back propagation, gradient aggregation and (global) weight updating. First, the forward propagation performs a series of linear or nonlinear operations given the input data for every layer from the first to last.
A layer's output is the input of the next layer. Then the observed output is compared with the expected and a loss value is calculated from the difference. Second, the backward propagation runs through from the last layer to the first by feeding the difference to the network and computes the gradient of the parameters.
Last, we update the weights with the gradients based on SGD. These three stages are repeated many times during a training.

It would be extremely expensive for the computation to update the weights with a large-scale training set at one time. Bottou developed a \emph{mini-batch SGD}~\cite{bottou2010large} approach to solve the slow weight update process. A training data set contains a number of data samples. The mini-batch SGD shuffles all samples and groups them into mini-batches. It employs a number of workers to work on these mini-batches in parallel. For a single mini-batch of samples, a worker performs forward propagation, backward propagation, and then computes the average gradient locally. Then global averaging is done and therefore the weights are updated per worker.

Distributed training is parallelized across a great number of workers. Fig.~\ref{fig:timeline} shows a typical process of distributed training of a neural network. Each worker owns a copy of the network model and hence a copy of the weights. The initial weights for each worker are usually randomly generated. Afterwards, a mini-batch is sent to each worker in parallel (not shown in this figure). All workers execute forward propagation to compute loss and backward propagation to compute gradients, and aggregate the gradients of a mini-batch locally. Then, due to the gradients of each worker are different, the gradients are averaged across different workers, which is in the form of Tree All-Reduce~\cite{agarwal2014reliable} or Butterfly All-Reduce~\cite{patarasuk2007bandwidth}. In Fig.~\ref{fig:timeline}, all gradients of different workers are averaged on worker 1. This process is divided into two steps: 1) the gradients of worker 3 and worker 4 are averaged to worker 3 and the gradients of worker 1 and worker 2 are averaged to worker 1; 2) the gradients of worker 1 and worker 3 are averaged to worker 1. Worker 1 updates the weights of model with these average gradients, and then broadcasts the updated model to all four workers again. Here, one iteration is  over.

 
\subsection{Communication Efficient SGD Algorithms}

\subsubsection{Gradient Compression and Sparsification}

Gradient sparsification \cite{lin2017deep,wangni2018gradient,alistarh2018convergence} and gradient quantization \cite{alistarh2017qsgd} focus on compressing gradients with efficient data representation and redundant communication elimination.
The default data format of gradients is single-precision floating-point $32$. Gradient quantization maps gradients from a format with the regular precision format to a format with lower precision or fewer bits~\cite{jia2018highly}, sometimes to ternary~\cite{wen2017terngrad} or binary~\cite{zhou2016dorefa, seide20141}. While quantizing gradients causes information loss, these works show that model converges with little accuracy loss. Deep Gradient Compression proposes momentum correction by accumulating quantization errors and using them at a later time. Gradient sparsification \cite{aji2017sparse, lin2017deep, renggli2019sparcml} explores that models are often over-parameterized and do not change all at once. Static or adaptive thresholds are used to determine significant gradients and are transferred for less communication bandwidth. These two groups of methods are orthogonal to our proposal. 

\subsubsection{Asynchronous SGD (ASGD)}
There are a few asynchronous training methods, such as Downpour SGD \cite{dean2012large}, Hogwild \cite{recht2011hogwild}, Elastic Averaging SGD\cite{zhang2015deep}. In these models, every worker has its own copy of weights. A worker performs forward propagation and back propagation on its own partition of samples, and then sends the calculated gradients asynchronously to a pool of parameter servers that manage a central copy of weights. The parameter servers update the central copy and then send the new weights asynchronously to each worker.
While each worker communicates gradients at a different time and avoids congestions at worker inter-connects, the parameter servers might be a performance bottleneck. 
For non-convex problems, ASGD requires that the staleness of gradients is bounded \cite{lian2015asynchronous} to match the convergence rate $ O (1 / \sqrt {K} )$ of synchronous SGD, where $K$ denotes the total Iteration steps.

\subsubsection{Local SGD}
Another set of methods targets at reducing the frequency of inter-worker communication and is called periodic averaging or Local SGD \cite{wang2018adaptive, wang2018cooperative,lin2018don}.
A worker performs SGD on its local copy of weights for $\tau$ times, where $\tau$ denotes the local iteration steps. After $\tau$ local updates, local copies are averaged across all workers globally in a synchronous manner. Several works suggested that Local SGD incurs the same convergence rate $ O (1 / \sqrt {K} )$ as SGD \cite{wang2018adaptive, wang2018cooperative}.
The total number of steps to train a model remains similar but the total amount of inter-worker communication is reduced by $\tau$ times. This has a similar effect as training with a large batch size, where the number of synchronizations decreases with an increase of batch size. However, a larger $\tau$ means more samples are processed on a single worker.
With Local SGD, SGD and forward/back propagations are still blocking while system resources are unbalanced.

%% file: theoretical.tex
\section{DaSGD}\label{section:theoretical}

\begin{figure}[t]
	\centering
	\includegraphics[width=\linewidth]{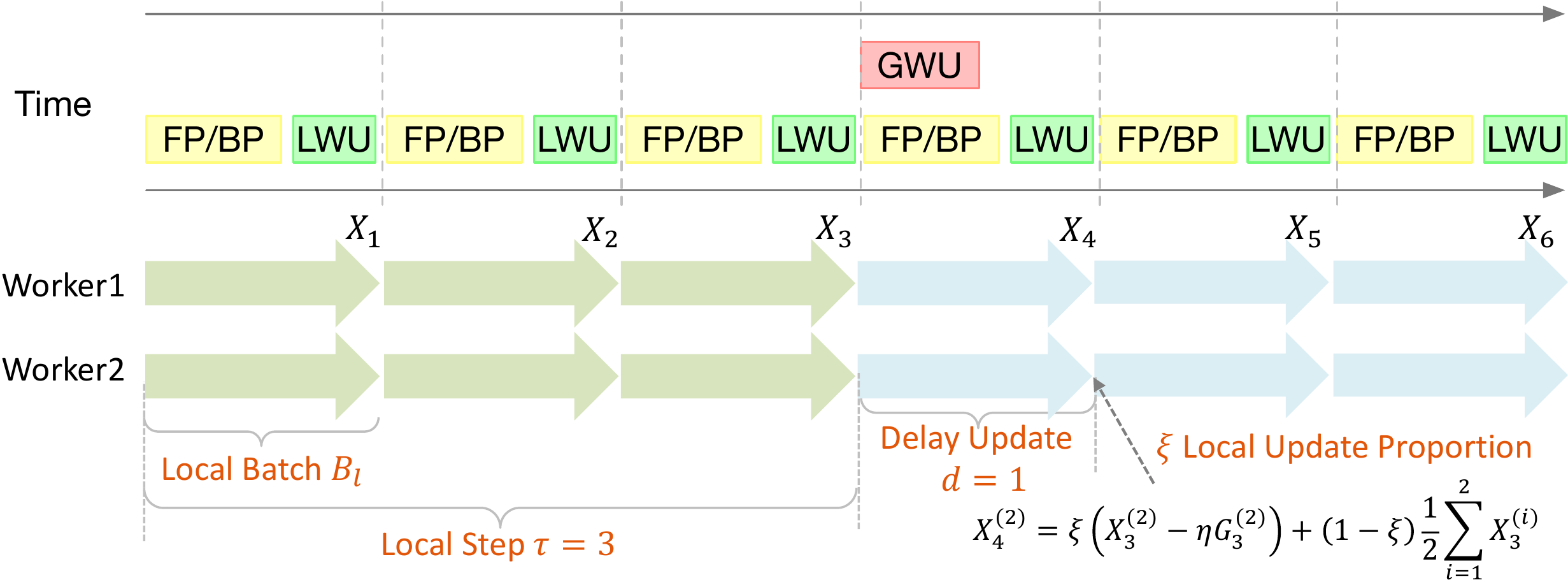}
	\caption{Timing Diagram of DaSGD. Six iterations of two workers are shown. Each arrow represents a local update with a local batch $B_l$, which includes a forward/backward propagation (FP/BP) and a local weight update (LWU). The number of local step $\tau$ is $3$, that is, after $3$ local updates, the global weight update (GWU) will occur. The updated global coped model is not update directly in the current iteration, but is updated on each local coped model proportionally after $d$ local updates, where the update proportion of local model is set as the auxiliary update parameter $\xi$.}
	\label{fig:structure}
\end{figure}

In this paper, we propose a new algorithm, called~\emph{Local SGD with Delayed Averaging}, ~\emph{DaSGD} for short. It aggregates gradients and updates weights in a relaxed manner, which helps parallelize the computation of forward/backward propagation with two other execution components:  the execution of global weight averaging and inter-worker data communication.

Our algorithm was initially inspired by the Local SGD algorithm~\cite{lin2018don, wang2018adaptive, wang2018cooperative} (discussed in Section~\ref{section:background}). Although Local SGD was designed to reduce communication and synchronization overhead~\cite{bottou2010large, dekel2012optimal}, it still involves a significant amount of communication overhead. To further decrease communication overhead, even to zero, the proposed algorithm exploits a delayed averaging approach 
that makes two novel improvements based on Local SGD. First, in order to merge remote weights by other workers with local in a deterministic way, DaSGD serializes forward propagations and back propagations for different samples. Second, workers start with local computations for the next samples while waiting for the aggregation and synchronization of global weights. In this way, the global communication and synchronization overhead is hidden or overlapped by local computations at the cost of a delayed update of local weights. However, theoretically we will prove that the convergence rate is the same as Mini-batch SGD. Furthermore, DaSGD parameterizes the overlapping degree so that when a large training cluster requires a longer time to synchronize, a worker may perform more iterations of local computations.

Fig.~\ref{fig:structure} illustrates the proposed algorithm by showing a wall-clock time diagram of $2$ training epochs. There are $2$ workers, dividing a global batch into 6 local batches. Each worker computes $3$ local batches. Each local batch contains $d$ samples. Each worker maintains a local copy of model. According to Local SGD, for a local batch, each worker operates $d$ forward/backward propagations and then updates the weights of its local model. After $2$ local updates, a worker synchronizes local weights with the other workers, resulting in an all-reduce operation being generated to average the  model weights. For example in Fig.~\ref{fig:structure}, all workers wait for, at local step $3$, the global synchronization to be finished and then start to operate on the next local batch each, in the scenario of Local SGD.

DaSGD implements a key feature by imposing delay update on Local SGD. As shown in Fig.~\ref{fig:structure}, a worker, at local step $3$, broadcasts its local weights to the wild and then immediately starts to compute on next local batch, without waiting for the global synchronization to be finished. Later, at local step $4$, the worker receives all the other workers' weights and then updates its local weights. This design very efficiently overlaps the communication of weights and forward/backward propagations of next local batch. 

In DaSGD, we use $\tau$ to denote the number of local batches between two consecutive global synchronizations. Therefore, $\tau$ is a controlling parameter that quantifies the number of propagations between weight averaging globally. During the delay update, both local computation and the global communication of weights are executed in parallel. As long as communication time is no more than the computation time of $d$ local iterations, the communication time can be hidden in the overall model training time. Careful tuning of $d$ and $\tau$ can realize full parallelism of global averaging and local computations. Unlike Local SGD, $\tau$ does not have to be large, as it is not only used to reduce inter-worker communication overhead~\cite{lin2017deep}.




In the following part of this section, in order to compare the proposed algorithm and traditional SGDs, we start with the update framework of each algorithm, and then qualitatively analyze execution time. Finally, we discussed the updated rules and the convergence rate in detail.

\subsection{Update Flow of Different SGD}
\begin{figure}[t]
	\centering
	\includegraphics[width=\linewidth]{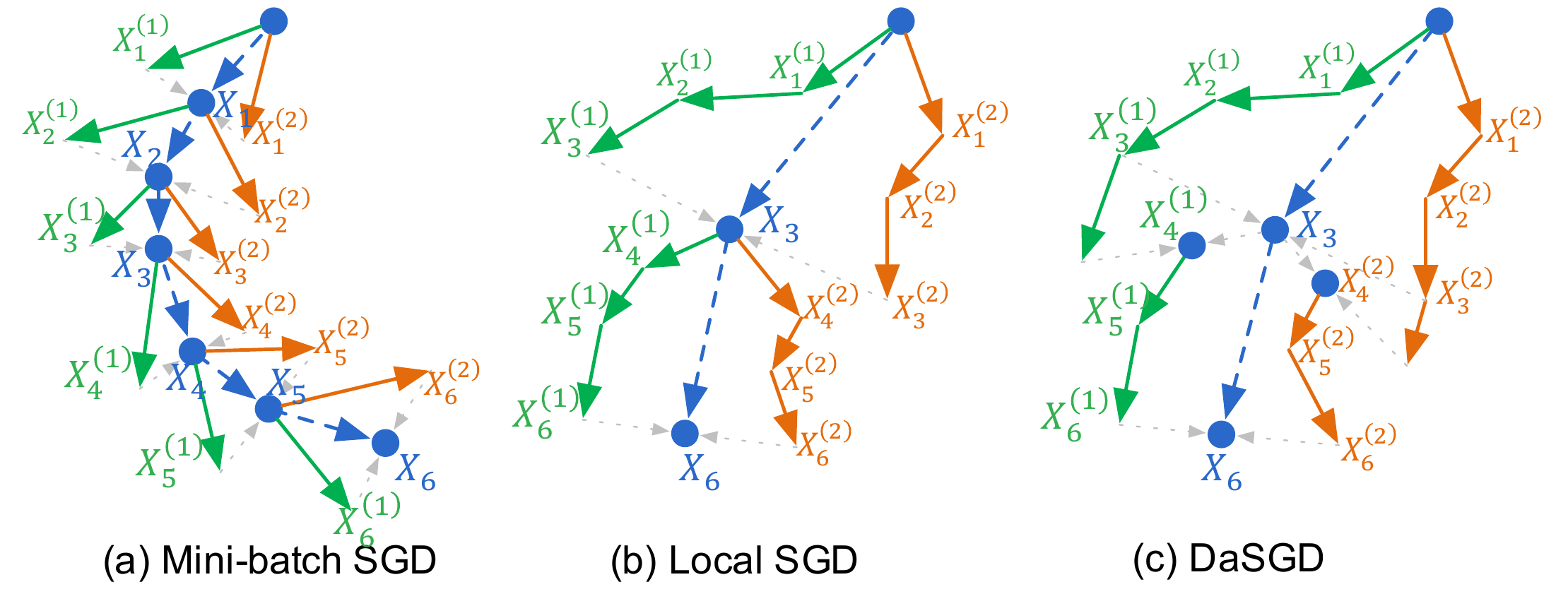}
	\caption{Loss landscape of (a) Mini-batch SGD, (b) Local SGD, and (c) DaSGD. 12 samples are updated on two workers. The orange and green arrows represent the updated loss function of each sample, and the blue arrows describe the location of the updated loss function on the global model.  }
	\label{fig:update_rule}\vspace{-3mm}
\end{figure}

Fig.~\ref{fig:update_rule} explains the mechanisms of weight update flows of~\emph{Mini-batch SGD}~\cite{bottou2010large,dekel2012optimal}, \emph{Local SGD}~\cite{lin2018don, wang2018cooperative, wang2018adaptive}, and DaSGD by taking an example of a $2$-worker parallel training process that sets the batch size as $2$ samples. The $2$ workers are distinguished by yellow and green arrows. In the Mini-batch SGD (as shown in Fig.~\ref{fig:update_rule}(a)), every worker updates its local weights once every \emph{mini-batch}, which is computed as the batch size divided by the number of workers. When both workers finish local updates for a mini-batch, local weights are merged to compute their average (shown by blue arrows). Next, both workers update their local weights with the average. Local SGD (shown in Fig.~\ref{fig:update_rule}(b)) reduces the weight aggregation times by letting every worker first update weights locally for continuous $\tau$ \emph{local batch}es in a row before a global merge is made. Local batch in the context of Local SGD is just a synonyms of the mini-batch in the context of Mini-batch SGD.

Same as the regular periodic averaging method (i.e., Local SGD), in the proposed algorithm, each worker updates local weights for $\tau$ local batches before a global aggregation. A novel change made by the proposed algorithm is to delay weight update from global to local after the global averaging. A worker may delay the update for $d$ \emph{step}s (i.e., samples) of local weight updates ($d = 1$ in this example, as shown in Fig.~\ref{fig:update_rule}(c)). With this novel algorithmic design, the time of global weight averaging can be hidden by parallelizing it with local computation by a worker, i.e., forward propagation, backward propagation, and local weight update. Large $d$ can be set if the time of global weight aggregation is very long in a large-scale distributed training system to shorten the overall training time.

\begin{figure}[t]
	\centering
	\includegraphics[width=\linewidth]{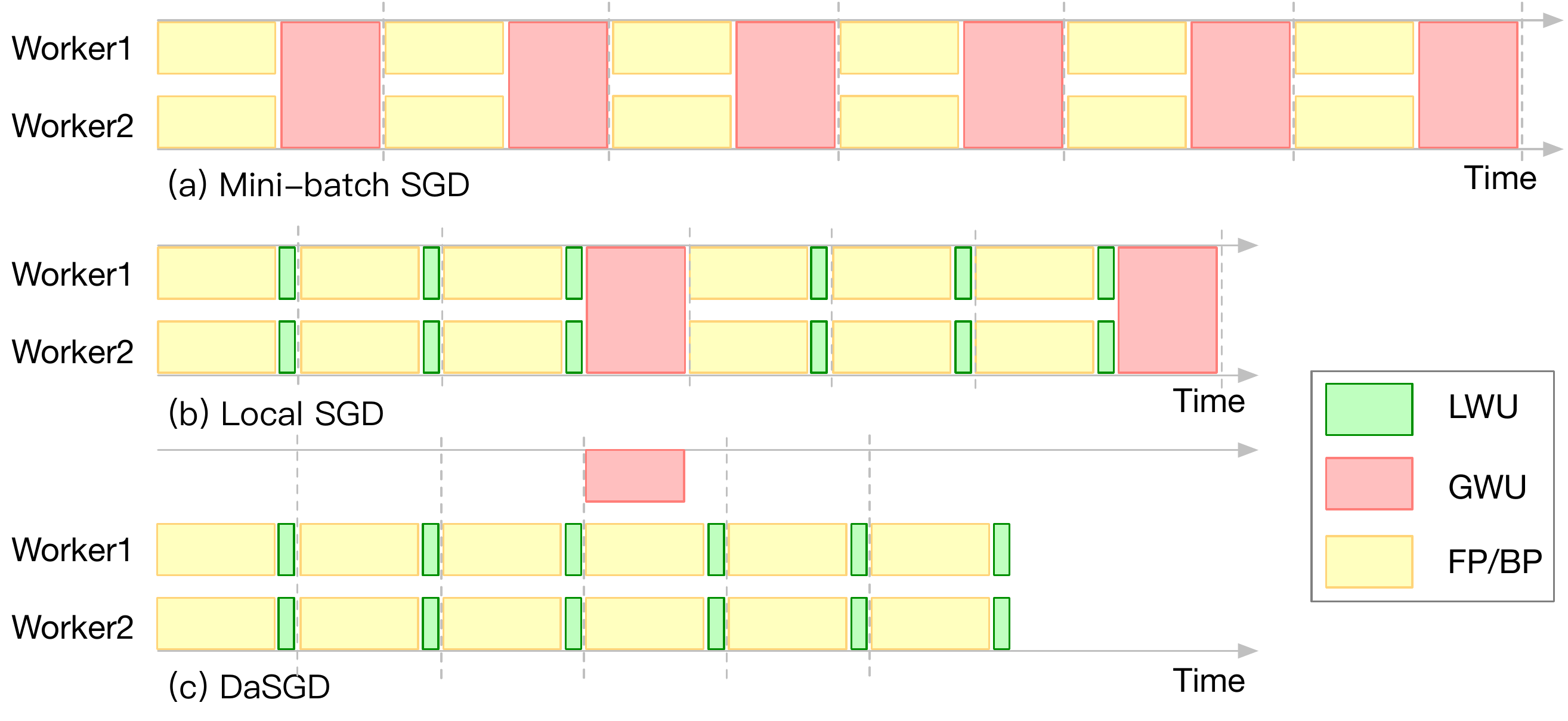}
	\caption{Execution time diagrams of different SGDs.}
	\label{fig:3compare}
\end{figure}

\subsection{Execution Time}
Before discussing the convergence rate, we first qualitatively analyze and compare the execution time between \emph{Mini-batch SGD}, \emph{Local SGD} and DaSGD. Figure~\ref{fig:3compare} presents schematic diagrams of the three SGD  algorithms for $6$ iterations. For Minibatch SGD, the weights are aggregated after every iteration, so the total execution time is measured by $6$ communications and $6$ local computations. By setting $\tau$ as 3, Local SGD reduces to $2$ communications, with the total execution time measured by $2$ communications and $6$ computations. Expectedly, DaSGD performs the best by hiding communication time cost in the delayed weight update. As a result, the total execution time of DaSGD is measured by just $6$ computations.

\subsection{Convergence Analysis}
\subsubsection{Update rule}
The update rule of our algorithm is given by
\begin{equation}\label{equ:updaterule}
\begin{aligned}
x_{k+1}^{(m)}  = & \left \{
\begin{array}{lll}
x_{k}^{(m)} - \eta g \left(  x_{k}^{(m)} \right)   , \quad\quad  {\rm otherwise} \\
\xi x_{k}^{(m)} - \eta\xi g \left(  x_{k}^{(m)} \right)
+\frac{(1-\xi)  \sum\limits_{j=1}^{M}\left[  x_{k-d}^{(j)} - \eta g \left(  x_{k-d}^{(j)} \right)  \right] }{M} \\ 
~~~~~~~~~~~~~~~~~~~~~~,\quad \quad (k+1)\mod \tau = d
    \\
\end{array}
\right.\\
\end{aligned}
\end{equation}

\noindent
where $x_{k}^{(m)}$ is the weights of worker $m$ at $k$-th iteration, $\eta$ the learning rate, $M$ the number of workers, and $g(x_{k}^{(m)} )$ the stochastic gradient of worker $m$. For every $k$ that satisfies $(k+1)~{\rm mod}~\tau = d $, a global average is updated to local weights. Besides, $\xi$ is an auxiliary parameter to adjust the weight of local weights in contrast to the global average when fusing them together.

We define the average weight and the average gradient
$$ \mu_{k}=\frac{1}{M} \sum_{i=1}^{M}   x_{ k}^{(i)} ,
~\bar{g}_k =\frac{1}{M} \sum_{i=1}^{M} g \left(  x_{k}^{(i)} \right).$$

After rearranging, the update rule for the average weight is obtained by
$$\mu_{\tau (k+1)+d} 
= \mu_{\tau k+d}-\eta \left[ \xi   \sum\limits_{i=\tau-d}^{\tau-1} \bar{g}_{\tau k+d+i}
+ \sum\limits_{i=0}^{\tau-1-d}  \bar{g}_{\tau k+d+i} \right] $$
It is observed that the averaged weight $\mu_{\tau (k+1)+d}$ is performing a perturbed stochastic gradient descent. Thus, we will focus on the convergence of the averaged weight $\mu_{\tau (k+1)+d}$,  which is common approach in the literature of distributed optimization \cite{wang2018cooperative,wang2018adaptive}.
SGD can converge to a local minimum or saddle point due to the non-convex objective function  $F(x)$. Therefore, the expected gradient norm is used as an index of convergence.

\subsubsection{Assumptions}
The common assumptions of the SGD analysis are defined as the following constraints \cite{wang2018cooperative}: 
\begin{itemize}
	\item  Lipschitzian gradient: $ ||\bigtriangledown F(x)- \bigtriangledown F(y) || \leq  L||x-y||$
	\item Unbiased gradients: $ E_{\mathcal{S}_k | x}\left[ g(x)\right]=   \bigtriangledown F(x)$	
	\item Lower bounder: $ F(x) \geq F_{inf} $		
	\item Bounded variance: $  E_{\mathcal{S}_k| x} ||g(x)-\bigtriangledown F(x)  ||^2 \leq \beta  ||\bigtriangledown F(x)||^2+\sigma^2 $
	\item Independence: All random variables are independent to each other
	\item Bounded age: The delay is bounded, $d \leq \tau$
\end{itemize}
where $\mathcal{S}$ is the training dataset, $\mathcal{S}_k$ is set $\left\lbrace s_{k}^{(1)},...,s_{k}^{(M)} \right\rbrace $ of randomly sampled local batches, $L$ is the Lipschitz constant.

\subsubsection{Convergence Rate}
The learning rate is usually set as a constant and is decayed only whenthe training process is saturated. Therefore, we analyze the case of fixed learning rate and study the lower limit of error at convergence.

\textbf{Theorem (Convergence of DaSGD).} 
Under assumptions, if the learning rate satisfies $
\eta  \leq \min \left\lbrace \sqrt{ a} , \sqrt{ b} \right\rbrace $, where $a$ and $b$ and shown in Appendix.
Then the average-squared gradient norm after $K$ iterations is bounded as follows
\begin{equation}\begin{aligned}\nonumber
&\mathbb{E}\left[ \frac{1}{K}\sum\limits_{k=1}^{K}  
\left\|  \bigtriangledown F(\mu_{k})\right\|^2 \right] \\
\leq &\frac{2M\left[ F(\mu_1) - F_{inf}\right]   
	+   2MKL\eta^2\sigma^2 \left[ \xi^2 d + \tau-d \right]   }{\eta MK(\xi d+\tau-d)} \\
&+\frac{3\eta^4 \xi L^2  (\tau-d+d\xi )   }{MK(\xi d+\tau-d )} \frac{\xi^{2}}{1-\xi^{2}}   \left\| \sum\limits_{i=1}^{d-1} g(\textbf{X}_{d-1}) \right\|^2_F  \\
&+ \frac{6 \eta^4 L^2 \sigma^2  }{\xi d+\tau-d  }   \left( 
\frac{\tau\xi^{2}(\tau-d+\xi d)}{1-\xi^{2}}
+(\tau-d)^2+\xi d(\tau -1)
\right)
\end{aligned}\end{equation}
where $\boldsymbol{X}_k = \left[x_{k}^{1}, ... , x_{k}^{m}\right] $, $\left\|~\right\|_F^2 $ is the Frobenius norm.All proofs are provided in the Appendix.

\textbf{Corollary.} 
Under sssumptions, if the learning rate is $\eta= A/\sqrt{K}$ the average-squared gradient norm after $K$ iterations is bounded by
\begin{equation}\begin{aligned}\nonumber
&\mathbb{E}\left[ \frac{1}{K}\sum\limits_{k=1}^{K}  
\left\|  \bigtriangledown F(\mu_{k})\right\|^2 \right] \\
\leq &\frac{2M\left[ F(\mu_1) - F_{inf}\right]   
	+   2MLA^2\sigma^2 \left[ \xi^2 d + \tau-d \right]   }{AM\sqrt{K}(\xi d+\tau-d)} \\ +&\frac{3A^4 \xi L^2  (\tau-d+d\xi )   }{MK^3(\xi d+\tau-d )} \frac{\xi^{2}}{1-\xi^{2}}   \left\| \sum\limits_{i=1}^{d-1} g(\textbf{X}_{d-1}) \right\|^2_F  \\
+ &\frac{6 A^4 L^2 \sigma^2  }{K^2(\xi d+\tau-d)  }   \left[
\frac{\tau\xi^{2}(\tau-d+\xi d)}{1-\xi^{2}}
+(\tau-d)^2+\xi d(\tau -1)
\right] 
\end{aligned}\end{equation}

If the total iterations $K$ is sufficiently large, then the average-squared gradient norm will be bounded by 
\begin{equation}\begin{aligned}\nonumber
&\mathbb{E}\left[ \frac{1}{K}\sum\limits_{k=1}^{K}  
\left\|  \bigtriangledown F(\mu_{k})\right\|^2 \right] \\
\leq&
\frac{2M\left[ F(\mu_1) - F_{inf}\right]   
	+   2MLA^2\sigma^2 \left[ \xi^2 d + \tau-d \right]   }{AM\sqrt{K}(\xi d+\tau-d)} 
\end{aligned}\end{equation}

Therefore, on non-convex objectives, the convergence rate of the proposed algorithm is consistent with the Mini-batch SGD and the Local SGD as $ O (1 / \sqrt {K} )$.

\subsection{Guidelines for Using DaSGD}
DaSGD is similar to Local SGD, the only difference is that the global model is updated to every local workers after $d$ local steps. The adjustment of other parameters is the same as that of Local SGD. Here we mainly discuss the setting of delay, which is the key of DaSGD. In order to realize the parallel communication and computation in DaSGD, the weight/gradient transfer time $t_c$ across workers is required to be less than $d$ local iteration time, that is, $t_c < d t_p$, where $t_p$ is the computation time in one local update. For deep learning systems, the weight/gradient transfer time $t_c$ across multiple workers among multiple is approximately calculated as the number of parameters $n_p$ of neural network models multiplied by the number of workers $m$ divided by bandwidth $\text{BW}$ of the device, $t_c = m n_p /\text{BW}$. The computation time $t_p$ in one local update is approximately calculated as the FLOP (floating-point operation) counts of the operation multiplied by local batch size divided by the computation speed FLOPS (floating-point operation per second) of the device, $t_p = B_l \text{FLOP}/\text{FLOPS}$. Therefore, the delay is given by 
\begin{equation}d>\frac{t_c}{t_p} = \frac{m\cdot n_p\cdot\text{FLOPS}}{B_l\cdot\text{BW}\cdot\text{FLOP}}.\end{equation}
It is worth noting that the delay is related to the structure of neural network models (the number of parameters and FLOP) and the configurations of deep learning systems (the local batch, the worker number, the bandwidth of the device and the computation speed). The current deep learning system has significantly improved the bandwidth and performance, and the discovery of residual network makes the growth of network parameters not obvious. So in most cases, when the delay is $1$, the weight/gradient transfer can be processed completely in parallel with local updates. In addition, as the worker number increases, the increase of the worker number will lead to the increase of the number of the transferred weight/gradient increases, and the delay needs to be increased moderately. The cooperative design of various parameters in DaSGD and hardware is discussed in detail in the following sessions.

%% file: experimental.tex
\begin{table}[t]
	\caption{Accuracy of DaSGD, Mini-batch SGD and Local SGD on CIFAR-10. }
	\begin{center}
		\vspace{-2mm}
		{
			\begin{tabular}{l|c| c|c  }
				\hline
				\hline
				\multirow{2}*{Model}  		& \multicolumn{3}{c}{Accuracy after 50 epochs}   \\
				\cline{2-4}
				& Mini-batch SGD  & Local SGD  & DaSDG  \\
				\hline
				GoogleNet      & $0.9409$ & $0.9468$ & $0.9444$\\
				VGG-16      	 & $0.9264$ & $0.9330$ & $0.9343$\\
				ResNet-50      & $0.9037$ & $0.9062$ & $0.9088$\\   
				ResNet-101     & $0.9019$ & $0.9061$ & $0.9045$\\
				DenseNet-121 & $0.9332$ & $0.9369$ & $0.9357$\\
				MobileNetV2   & $0.9304$ & $0.9241$ & $0.9304$\\
				ResNeXt29     & $0.9403$ & $0.9424$ & $0.9415$\\
				DPN-92      	& $0.9354$ & $0.9513$ & $0.9502$\\
				\hline
				\hline
		\end{tabular}}	\label{table:3compareaccuracy} 
	\end{center}
\end{table}
\begin{figure*}[t]
	\centering
	\includegraphics[width=\linewidth]{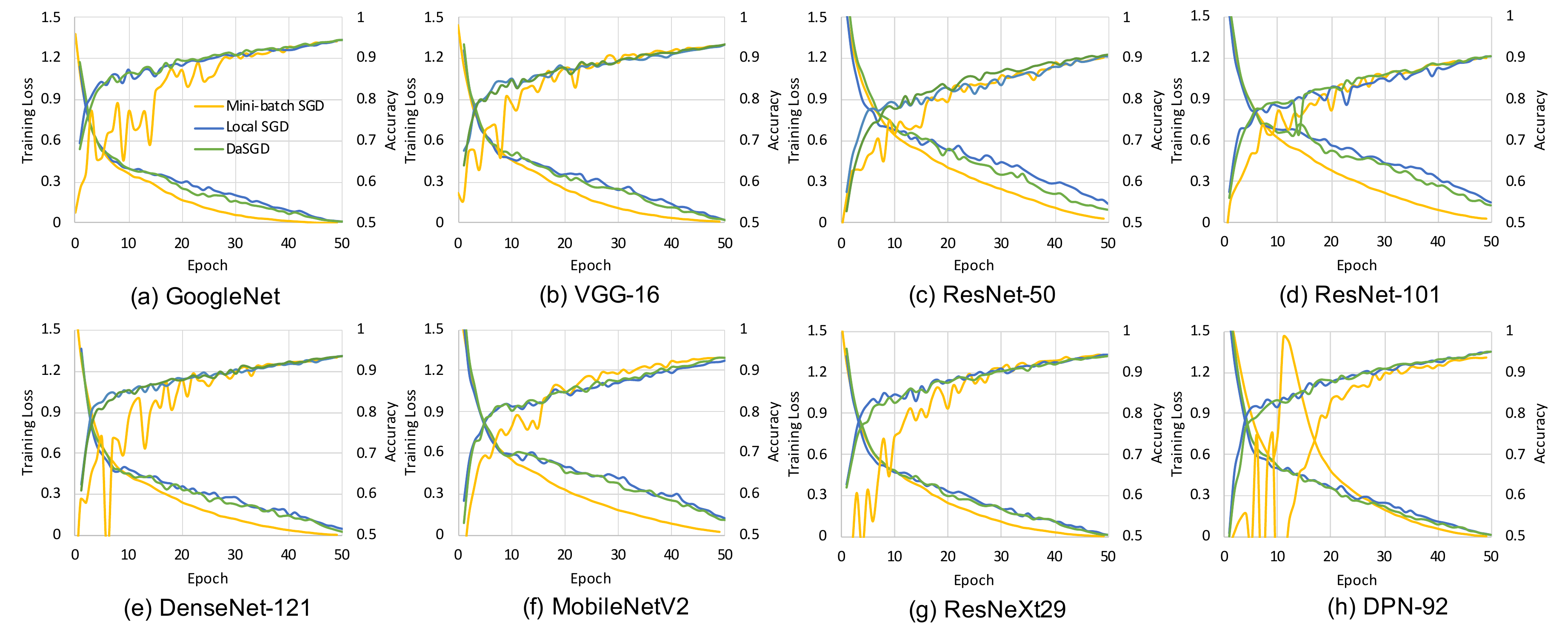}
	\caption{Training loss and accuracy in different model based on CIFAR-10 dataset.}
	\label{fig:3sgd}
\end{figure*}

\section{Experimental Results}\label{section:experimental}
In this session, we will introduce our experimental settings and the convergence rate of DaSGD, Local SGD and Mini-batch SGD for different models. Then the influence of DaSGD parameters on the convergence rate in given. 
\subsection{Parameter Setup}
The training process is implemented under the Fast.Ai \cite{fastai} platform based on CIFAR-10 dataset. The learning rate is adopted \textit{One Cycle Policy} \cite{smith2017cyclical}, which makes it linearly increase first (from $0.0001$ to $0.01$ in $30\%$ epochs) and then linearly decrease (from $0.01$ to $0.0001$ in $70\%$ epochs) within a reasonable range. A higher learning rate helps to prevent the model from falling in the steep area of the loss function, hoping to find a flatter minimum; A lower learning rate prevents training from diverging and converging to a local minimum. This learning schedule improves the accuracy in fewer iterations, allowing us to get more accurate results in only $50$ epochs.
The weight decay is $0.01$ and the moment is $0.9$.
Since we only want to analyze the convergence rate and accuracy, the comparison with the Local SGD and Mini-batch SGD is performed in $50$ epochs.

\subsection{Convergence Rate and Accuracy}
Compared with the Mini-batch SGD and Local SGD in distributed training, we analyze the convergence rate and accuracy. TABLE~\ref{table:3compareaccuracy} shows the accuracy of Mini-batch SGD, Local SGD and DaSGD after $50$ epochs based on CIFAR-10 dataset. It includes the existing common neural network models, such as GoogleNet \cite{googlenet},  VGG-16\cite{vgg}, ResNet-50 \cite{resnet}, ResNet-101, DenseNet-121 \cite{densenet}, MobileNetV2\cite{mobilenetv2}, ResNeXt29\cite{resnext}, and DPN-92\cite{dpn}. All models are trained under $32$ workers. The total batch size of Mini-batch SGD is $1024$. According to the data parallelism, the batch size distributed to each worker is $32$. The local batch size $B_l$ of Local SGD and DaSGD is $32$, the number of local steps $\tau$ is $4$, and the delayed iteration steps $d$ of DaSGD is $1$. For the three algorithms, the total number of iterations is the same under different models, which is $2450$.

As shown in TABLE~\ref{table:3compareaccuracy}, we can find that for different models, with $1$K batch size, the network model with higher accuracy can be obtained in a short iteration steps without adjusting the hyper-parameters. 
Due to the large batch size for each iteration of Mini-batch SGD, the hyper-parameters needs to be adjusted carefully. The optimization difficulty leads to the accuracy loss for large-batch training.  Only the linear scale rule for adjusting the learning rate as a function of the total mini-batch size and the warm-up scheme are not enough. It is necessary to change the network structure, like adding batch normalization, for the high-accuracy training. These additional optimization methods for large-batch training are complex and tedious, and the algorithm based on local update overcomes this problem since the batch size of local updates is small. Thus, without any hyper-parameter adjustment for large-batch training, in addition to MobileNetV2, the accuracy of Local SGD and DaSGD is higher than that of the Mini-batch SGD. 
Fig.~\ref{fig:3sgd} shows this more clearly. At the beginning of distributed training, since the batch size is large, the algorithm based on Mini-batch SGD is usually very unstable, and the accuracy fluctuates greatly. The convergence rate is slower than that of the Local SGD and DaSGD. At the end of  training, although the training loss of Mini-batch SGD is smaller, Local SGD and DaSGD has small test loss and higher accuracy.

\subsection{Parameter Influence of DaSGD}
We evaluate the influence of different parameters on the convergence rate and accuracy in ResNet-50 model.  Five adjustable parameters in DaSGD algorithm, which are the number of workers, the local batch size, the number of local step, the local update proportion and the delay, are discussed and analyzed respectively in Fig.~\ref{fig:resnet50fig}.
The baseline is set, where the number of workers $m$ is $32$, the local step $\tau$ is $4$, the delay is $2$, the local batch $B_l$ is $32$, the local update proportion $\xi$ is $0.25$. 

\subsubsection{Worker number}
Fig.~\ref{fig:resnet50fig}(a) shows the accuracy of different worker numbers based on ResNet-50, illuminating that DaSGD has a fast convergence rate and high accuracy in general. As the number of workers increases from $2$ to $256$, the convergence rate slows down and the accuracy decreases. Since the local batch size remains unchanged as $32$, when the worker number is $256$, the total batch size has reached $8192$, resulting in a decrease of accuracy of about $2\%$ and a high training loss. In addition, DaSGD only communicates across workers every four local updates, and the samples of four local iterations has reached $32$k, which is a huge batch for CIFAR-10 dataset with only $50000$ training samples. 
The effect of the worker number on the distributed training is mainly reflected in that increasing the worker number can accelerate the training process, but the increase of the worker number leads to the linear increase of weight/gradient transmission, which increases the communication time and weaks the acceleration. Since forward/backward propagation and weight/gradient transfer are parallel, the increase in communication time caused by the increase of worker number is not reflected in the total execution time.
However, in order to eliminate the increase of communication time driven by the increase of worker number in parallel, it is necessary to increase the delay update steps appropriately when the number of workers increases to a certain extent. This part is discussed in detail in Session \ref{section:strategy}. Through the analysis system model, we can evaluate the communication time  under multiple workers and computation time of one local update, and determine the number of delay update steps to make the communication process completely parallel.

\subsubsection{Local batch size}
Fig.~\ref{fig:resnet50fig}(b) illustrates that the DaSGD algorithm has a poor convergence rate for too large or too small local batch size. When the local batch is too large as $256$, the accuracy is significantly reduced, and when the local batch is too small as $8$, the convergence rate is slowed down.
This phenomenon also exists in the Mini-batch SGD. Too large batch size leads to poor generalization ability, but it can reduce the total number of iterations; while too small batch size reduces the generalization error due to noise, but it requires a large number of iterations.
Therefore, the selection of the local batch size is very important for DaSGD. Fig.~\ref{fig:resnet50fig}(b) demonstrates that the local batch size of $32$ or $64$ has high accuracy and low training loss.
It is worth noting that the total batch size is described as $B=m B_{l}$. When the worker number is $32$ and the local batch size is $256$, the total batch size rises to $32$k, which is faced with the problem of adjusting hyper-parameter of large-batch training discussed above. The convergence rate of training needs more cooperation with the adjustment of hyper-parameters at such a high batch size.

\subsubsection{Local step}
When the number of local steps increases from $4$ to $32$, the accuracy of DaSGD decreases slightly and the training loss increases, as shown in Fig.~\ref{fig:resnet50fig}(c). For DaSGD algorithm, the number of local steps should be reduced as much as possible under the condition of ensuring parallel communication, which is very different from Local SGD. 
By increasing the number of local steps, the Local SGD allocates time to several local iterations, resulting in a reduction in total execution time. In other words, increasing the number of local steps increases local iterations, which reduces the frequency of weight/gradient transfer across different workers. In order to reduce communication time, a large local step is required in Local SGD to share the communication time at the cost of accuracy loss. In addition, communication time is not essentially eliminated. Local SGD realizes the trade-off between communication time and accuracy by using local steps.
While, DaSGD only uses the local step as a quantitative method to describe parallel communication. As long as the local step computation time is larger than the weight/gradient communication time, the communication time can be eliminated in the total execution time. Therefore, the DaSGD algorithm requires a small number of local steps, which is conducive to convergence rate and high accuracy.

\subsubsection{Update proportion}
\begin{figure*}[t]
	\centering
	\includegraphics[width=\linewidth]{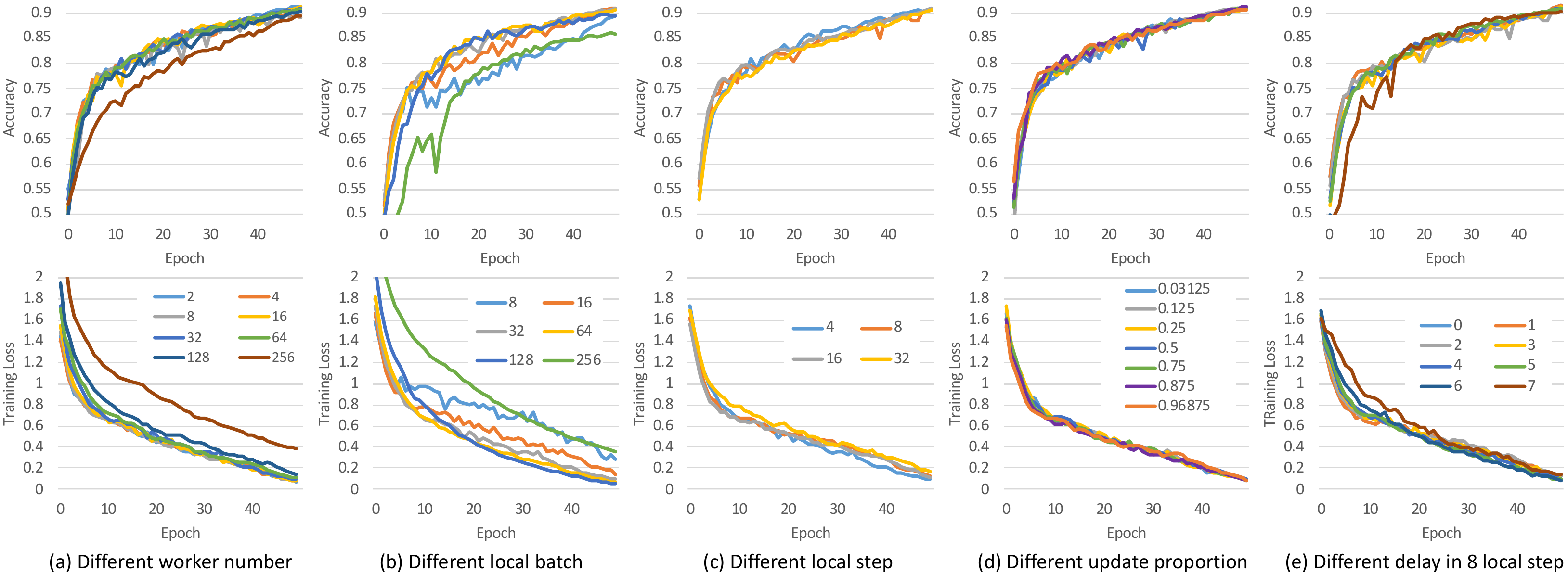}
	\caption{Effect of different parameters of DaSGD based on ResNet-50.}
	\label{fig:resnet50fig}
\end{figure*}

Fig.~\ref{fig:resnet50fig}(d) shows that the different proportions of local weights in the delay update of global weights have little effect on accuracy. From the update rule \eqref{equ:updaterule},  the local update proportion has the same meaning as the momentum in hyper-parameters. One cycle policy in Fast.Ai has shown that different momentum has little effect on accuracy.
\subsubsection{Delay}
The difference between DaSGD and Local SGD is that DaSGD delays the average model of every $\tau$ local steps by $d$ local update steps. The number of delay is closely related to the number of local steps. Fig.~\ref{fig:resnet50fig}(e) and (c) shows the two relationships between the number of delay update and the number of local steps, in which one is to keep the number of local steps and change the delay and the other is to keep the number of delay and change the local steps. In addition, the delay update is also limited by the local step.  It is assumed that it is smaller than the local step, that is, the global model update of the current iteration must be completed before the next global update. 
Delay has little effect on the convergence rate in general. When the delay increases from $0$ to $7$, the convergence rate slows down and the accuracy decreases, as shown in Fig.~\ref{fig:resnet50fig}(e). The Local SGD is shown as the delay is $0$, so the accuracy of DaSGD is slightly lower than that of Local SGD in the same local steps.
Besides, a large delay update is usually not implemented. Since the weight/gradient transfer time is relatively small compared to the forward/backward propagation time of the local iteration, $1$ delay update can eliminate the weight/gradient communication time in the total execution time in most cases. Due to the increase in the worker number, the time of the weight/gradient transfer across workers may be longer than the forward/backward propagation time of local iterations. In this case, the number of delay update can be appropriately increased to eliminate communication time, which is also discussed in the influence of the worker number part.

%% file: system.tex
\section{System Performance Evaluation}\label{section:system}

\subsection{Analytical Model of Distributed Training Performance}
We analyze the performance of the distributed system under different SGD algorithms and show the analytical model.
The performance of real-life systems are affected by many issues in the system hardware and software stacks, such as whether the software framework overlaps communication and computation, what reduction algorithm is used, how GPUs interconnects, how much network throughputs the servers have. The differentiation of an algorithm may be obscured by these issues.
We abstract an analytical model with the following key performance parameters based on the system configuration and the training setup. 
The total execution time $t_{total}$ of distributed training is decomposed into forward propagation time for a single sample $t_{f}$, backward propagation time for a single sample $t_{b}$, the time for gradient aggregation and weight update on the same worker $t_{l}$, the time for gradient aggregation and weight update among multiple workers that are not hidden behind computation time (communication time) $t_{c}$. The total amount of training data in a dataset is defined as $n_s$, the number of samples a worker computes parallel is defined as $p$, and the number of workers is defined as $m$.
\begin{figure*}[t]
	\centering
	\includegraphics[width=\linewidth]{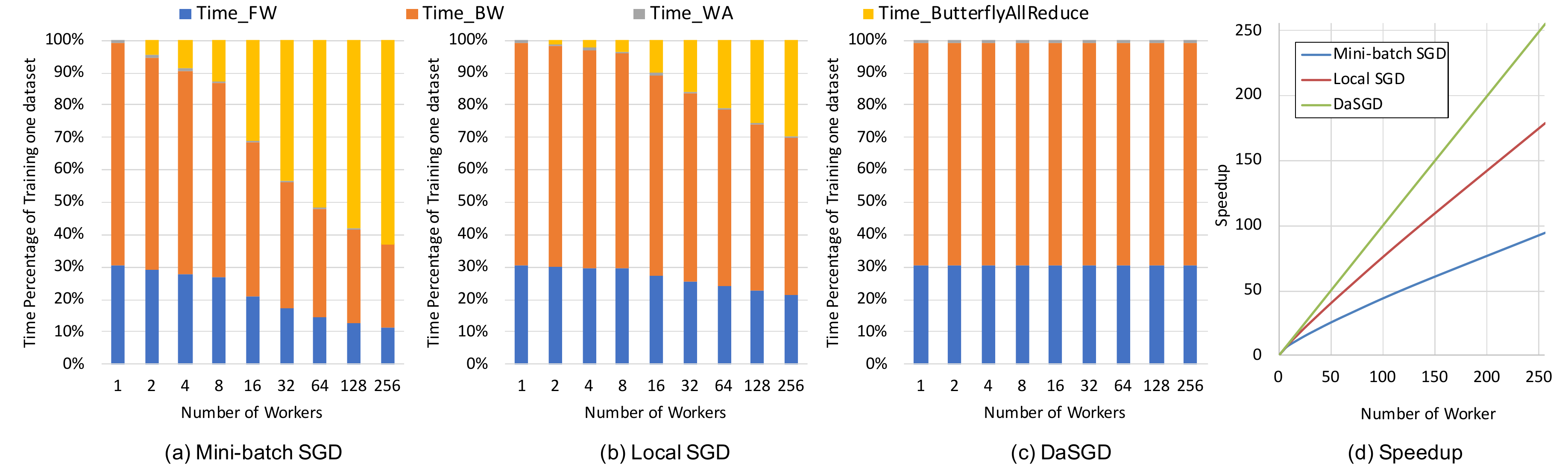}
	\caption{PALEO analytical results of data parallel distributed training of ResNet-50 with up to $256$ servers on the NVIDIA TITAN X GPUs. A comparison of (a) Mini-batch SGD, (b) Local SGD, and (c) DaSGD is shown. (d) Weak scaling speedup results based on the Butterfly AllReduce communication scheme. }
	\label{fig:paleo}
\end{figure*}
\subsubsection{Mini-batch SGD}
We formulate the training process using Mini-batch SGD into three steps. 
(a). Forward/backward propagation. Forward propagation is performed layer by layer in each worker, and the gradient is generated using the chain rule to realize the backward propagation. The samples in a mini-batch are divided to $m$ worker and are processed in $p$-parallel. Each worker performs $B/pm$ times forward/backward propagations in a mini-batch.
(b). Local gradient accumulation on each local worker.  We assume the framework is optimized and gradients from samples are accumualted locally at the worker first before being synchronized among workers. Each worker perform local gradient accumulation for $t_l$ in a mini-batch.
(c). Gradient aggregation and weight update among all workers. Each worker needs $t_c$ for this in a mini-batch.
Therefore, the total time of training a model $t_{total}$ is described as
\begin{equation}
t_{total} =\left[  \frac{B}{pm}   \left(t_{f}+t_{b}   \right)   + t_l +t_c          \right]  \frac{n_s}{B}.
\end{equation}

\subsubsection{Local SGD}
Local SGD needs to complete $\tau$ local updates before global model averaging. We formulate it into the the following steps: $\tau$ 
(a). local updates: each worker completes $\tau B/pm$ forward and back propagation.
(b).  $\tau$ local SGD aggregation and local weight update. 
(c). Update model every $\tau$ local updates using the average local model between all workers.
The  total execution time is represented as
\begin{equation}
t_{total} =\left[  \frac{B}{pm}   \left(t_{f}+t_{b}   \right)   + t_l +\frac{t_c }{\tau}       \right]  \frac{n_s}{B}.
\end{equation}
The above formula also proves that the difference between Local SGD and Mini-batch SGD is that the communication time $t_c  $ in Mini-batch SGD is reduced $\tau$ times to $t_c/\tau$. In order to reduce the weight/gradient transfer time effectively, a large $\tau$ value is usually required.

 \subsubsection{DaSGD}
 By delaying the local update of the global model, DaSGD algorithm realizes the local update computation and weight/gradient communication in parallel.  The process is similar to Local SGD involving three steps:
(a). $\tau$ local updates: each worker completes $\tau B/pm$ layer-by-layer feed forward calculation and gradient back propagation. 
(b). $\tau$ local SGD calculations include weight aggregation and weight apply.
(c). The averaging model is updated after $d$ local iterations every $\tau$ local updates .
The third step will not be reflected in the total execution time, since it can be performed in parallel with the previous two processes.
When $ t_c <d \left[ B\left(t_{f}+t_{b}   \right)/pm + t_l  \right] $, it means that the weight/gradient transfer time is shorter than that of $d$ local iterations, and the total execution time is showed as
\begin{equation}
t_{total} =\left[  \frac{B}{pm}   \left(t_{f}+t_{b}   \right)   + t_l     \right]  \frac{n_s}{B}.
\end{equation}
It is worth noting that compared with Mini-batch SGD and Local SGD, DaSGD completely eliminates communication time by parallelizing processing mode. The training process can be accelerated only by changing the algorithm without any special requirements for the deep learning system

\subsection{Performance Simulation}
In order to effectively evaluate the performance improvement of DaSGD to the distributed deep learning system for a given problem instance, we use the PALEO, a DNN performance model, which provides performance estimations within $10\% – 30\%$ prediction errors~\cite{qi17paleo}.
We analyze the distributed training of ResNet-50, which has $25.5$ million parameters and occupies $102$ MB of memory.
The experiments simulate a distributed training cluster with a less optimal configuration. It consists of up to $256$ the NVIDIA TITAN X GPUs with PCIe3.0. An enhancement is made on the Paleo to simulate the case each server uses PCIe3.0 ($16$ GBps) connecting $8$ GPUs with $20$ Gbps Ethernet between servers. 
The Butterfly AllReduce scheme is adopted for gradient aggregation. 

PALEO decomposes the total execution time $t_{total}$ into computation time and communication time with all layers included, which includes forward propagation time $t_{f}$, backward propagation time $t_{b}$, the time for gradient aggregation and weight update within a single worker $t_{l}$, and the time for gradient aggregation and weight update between workers (communication time) $t_{c}$.
Fig.~\ref{fig:paleo} shows a comparison of three algorithms for training the ResNet-50 with a mini-batch size $64$ on up to $256$ workers under weak scaling, where weak scaling means that the global batch size is increasing as increasing the number of workers. The execution time breakdown for various workloads of processing the whole dataset (one epoch) is shown in Fig.~\ref{fig:paleo}.  
\subsubsection{Gradient aggregation and weight update within a single worker}
The time for gradient aggregation and weight update within a single worker can be ignored, since it comprises a small percentage of the total training time, as shown in Fig.~\ref{fig:paleo}(a) and (b) (grey block).

\subsubsection{Forward/backward propagation}
The forward/backward propagations time of the three algorithms are the same (Fig.~\ref{fig:paleo}(a), (b) and (c)), since the batch size of each worker in each iteration is $64$. In addition, forward/backward propagations are compute-bound computations. Based on the PALEO performance model, the forward/backward propagation time of each layer of neural network is calculated in detail.

\begin{table*}[t]
	\caption{Parameters and time @256 workers, 64 local batch}
	\begin{center}
		\vspace{-2mm}
		{
			\begin{tabular}{l|r |c c c c c |c c c  cc}
				\hline
				\hline
				\multirow{3}*{Model} 		        &\multirow{3}*{Parameters}        
				& \multicolumn{5}{|c|}{TITAN X with $20$ Gbps Ethernet}  & \multicolumn{5}{|c}{K80 with $10$ Gbps Ethernet} \\
				\cline{3-12}
				&&\multirow{2}*{$t_p$}  & \multicolumn{2}{c}{$t_c$ AllReduce} & \multirow{2}*{delay}   & \multirow{2}*{$\tau$}  
				&\multirow{2}*{$t_p$}  & \multicolumn{2}{c}{$t_c$ AllReduce} & \multirow{2}*{delay}   & \multirow{2}*{$\tau$} \\
				&&		& Tree  &Butterfly  &  & &		& Tree  &Butterfly  &  &  \\
				\hline
				Network-in-Network & $7,595,176$ & $119.08$   & $132.91$ & $66.45$ & $2$ & $3$ 			& $129.80$ & $254.43$ & $127.21$ & $2$ & $3$\\
				VGG-16      	   & $138,357,544$ 		  & $2164.32$ & $2421.25$ & $1210.62$ & $2$ & $3$  & $2361.61$ & $4634.97$ & $2317.48$ & $2$ & $3$\\
				VGG-19 			   & $143,667,240$       & $2684.73$ & $2514.17$ & $1257.08$ & $1$ & $2$	& $2932.49$ & $4812.85$ & $2406.42$ & $2$ & $3$\\
				ResNet-50        & $25,530,472$  		& $526.05$   & $446.78$ & $223.39$ & $1$ & $2$		 & $575.29$   & $855.27$ & $427.63$ & $2$ & $3$\\ 
				ResNeXt-50      & $167,153,128$       & $1640.05$ & $2925.17$ & $1462.58$ & $2$ & $3$	& $1795.83$ & $5599.62$ & $2799.81$ & $4$ & $5$\\
				DenseNet-121   & $7,905,448$         & $358.23$   & $138.34$ & $69.17$ & $1$ & $2$ 			  & $390.73$   & $264.83$ & $132.41$ & $1$ & $2$\\
				DenseNet-201   & $17,900,106$ 		& $538.06$ & $313.25$ & $156.62$ & $1$ & $2$			& $587.64$ & $599.65$ & $299.82$ & $2$ & $3$\\
				\hline
				\hline
		\end{tabular}}	\label{table:system}\vspace{-5mm} 
	\end{center}
\end{table*}

\subsubsection{Gradient aggregation and weight update between workers}
In the Mini-batch SGD algorithm (Fig.~\ref{fig:paleo}(a)), when the number of workers is $256$, the gradient aggregation and weight update between workers contributes approximately $45.9\%$ of the total execution time.
It means that the larger the number of workers is, the more vulnerable the Mini-batch SGD algorithm is to be affected by the communication bottleneck.
When evaluating this overhead in Local SGD, it is reduced to $17.5\%$.  
The communication time of Local SGD is four times shorter than that of the Mini-batch SGD when the number of local update steps is $4$. The speedup of different algorithms with respect ot the number of workers is shown in Fig.~\ref{fig:paleo}(d).
A large number of workers does not scale at the linear rate of $1$, and the scaling rate is less for Mini-batch SGD, due to a larger proportion of time spent on gradient transfer.
In Mini-batch SGD, although increasing the number of workers can shorten the amount of computation time, it increases the total amount of data communication. The communication overhead increases linearly with the larger number of workers. 
Considering the high communication overhead for a large distributed training cluster, DaSGD parallels communication tasks with computation tasks and removes the weight/gradient transfer time from the total execution time.
As shown in Fig.~\ref{fig:paleo}(c), when DaSGD is applied, the time for gradient aggregation and weight update is mainly communication overhead and is completely hidden behind forward/back propagations.

%% file: strategy.tex
\section{Training System Design Strategies and Discussion}\label{section:strategy}
\subsection{Use the Hyper-Parameter Receipt for a Large Global Batch}
When the local batch size is $32$ and the worker number is $256$, the global batch size attains $8192$.  With regular Mini-batch SGD, training with a large batch size from $8$k to $64$k requires a specific set of hyper-parameter receipts. DaSGD needs no additional hyper-parameter adjustment to achieve high accuracy, but if it wants to achieve higher accuracy, it needs to further optimize the hyper-parameter receipt for large-batch training. This receipt is not used in our experiments.

\subsection{Select the Local Batch Size, Such as 32, 64}
Local batch size $32$ is a common practice for reasonable batch normalization results.
Local batch size affects accelerator performance. The sweet spot for a single GPU is at $128$ and $256$. But for better parallelism on a large cluster, $32$ or $64$ is recommended.

\subsection{Using System Analysis Model to Determine Delay}
DaSGD algorithm parallels the communication process with $d$-local updating, and requires that the time of weight/gradient transfer between workers is less than $d$-local update time, that is, $t_c < d t_p$. Weight/gradient transfer time can be calculated through the neural network model structure and the hardware parameters of the deep learning system, such as the network interconnection bandwidth, the number of workers, and the communication mode (Tree All-Reduce and Butterfly All-Reduce). While, the local update time included forward/backward propagation and weight aggregation/apply is determined by the number of parameters of neural network model and the peak FLOPS of deep learning system.
TABLE~\ref{table:system} analyzes the parameter number, the computation time of one local update $t_p$ and the weight/gradient communication time $t_c$ under the NVIDIA TITAN X GPUs system connected to $20$ Gbps Ethernet network and the GPU K80 system connected to a $10$ Gbps Ethernet network. For the deep learning system with high network interconnection bandwidth, the communication time is small. Even when the worker number is up to $256$ using the Tree AllReduce, the weight/gradient transfer can be completely parallel if the delay is $1$ or $2$. On the contrary, it is high in the Ethernet network with $10$ Gbps bandwidth.
In addition, Butterfly AllReduce optimizes that half of the nodes in Tree AllReduce do not send at the halving stage, and its communication time is about twice that of the Tree AllReduce. However, for large data block transferring, Butterfly AllReduce is prone to the communication time fluctuation caused by the insufficient utilization of bandwidth.

\subsection{Set Local Steps to Delay Plus One for High Accuracy}
The DaSGD algorithm requires that the global averaging model is updated after $d$ local steps. When it is updated to the local worker, the global averaging model is a stale calculation result. Updating the global averaging model to the local workers can effectively reduce the randomness between different local models, but the global model returned in the older version causes a slower convergence rate
. Therefore, we optimize the trade-off between the randomness in the local model and the staleness in the global model. In other words, we improve the trade-off between the number of local updates and the delayed update. The delay is obtained according to the model structure and the distributed system. Since the increase of the local steps reduces the accuracy, the number of local steps is set as the number of delay steps plus $1$, as shown in TABLE~\ref{table:system}, that is, $\tau = d +1$ to obtain higher accuracy and fast convergence rate. This means that the local step cannot be too long so that both the staleness of the delayed update and the randomness of the local models can be reduced.

%% file: conclusion.tex
\section{Conclusion}\label{section:conclusion}
In this work, we propose a new SGD algorithm called DaSGD, which parallelizes SGD and forward/back propagation to hide communication time.
Just adjusting the update schedule at the software level, DaSGD algorithm makes better use of distributed training systems and reduces the reliance on low latency and high peak throughput communication hardware.
Theoretical analysis and experimental results clarify that its convergence rate is $O\left( 1/\sqrt{K} \right) $, which is the same as the mini-batch SGD.
The auxiliary parameters are added to realize quantitative control, and their proper ranges and guidelines for using DaSGD are also provided.
The system evaluation demonstrates that DaSGD can speed up the deep learning system linearly without being weakened by high communication time.

%% file: appendix.tex
\onecolumn
\appendix
\section*{Convergence Analysis of DaSGD}

\subsection{Assumptions}
We define some notations. $\mathcal{S}$ is the training dataset, $\mathcal{S}_k$ is set $\left\lbrace s_{k}^{(1)},...,s_{k}^{(M)} \right\rbrace $ of randomly sampled local batches at $M$ workers in $k$ iteration, $L$ is the Lipschitz constant, $d$ is the number of local iteration that global weight updates are delayed, $\tau$ is the number of local steps, $x$ is the weight of devices. The convergence analysis is conducted under the following assumptions:

\begin{itemize}
	\item  Lipschitzian gradient: $ ||\bigtriangledown F(x)- \bigtriangledown F(y) || \leq  L||x-y||$
	\item Unbiased gradients: $ E_{\mathcal{S}_k | x}\left[ g(x)\right]=   \bigtriangledown F(x)$	
	\item Lower bounder: $ F(x) \geq F_{inf} $		
	\item Bounded variance in local SGD: $  E_{\mathcal{S}_k| x} ||g(x)-\bigtriangledown F(x)  ||^2 \leq \beta  ||\bigtriangledown F(x)||^2+\sigma^2 $
	\item Independence: All random variables are independent to each other
	\item Bounded age: The delay is bounded, $d \leq \tau$
\end{itemize}

\subsection{Update Rule}
The update rule of DaSGD is given by
\begin{equation}\nonumber
\begin{aligned}
x_{k+1}^{(m)}  = & \left \{
\begin{array}{ll}
x_{k}^{(m)} - \eta g \left(  x_{k}^{(m)} \right)   , \quad & {\rm otherwise} \\
\xi x_{k}^{(m)} - \eta\xi g \left(  x_{k}^{(m)} \right) +\frac{1-\xi}{M}\sum\limits_{j=1}^{M}\left[  x_{k-d}^{(j)} - \eta g \left(  x_{k-d}^{(j)} \right)  \right]     , &(k+1-d)\mod \tau = 0 \\
\end{array}
\right.\\
\end{aligned}
\end{equation}
where $x_{k}^{(m)}$ is the weights at $m$ worker in $k$ iteration, $\eta$ is the learning rate, $M$ is the number of workers, $g(x_{k}^{(m)} )$ is the stochastic gradient of worker $m$,  $\xi$ is the local update proportion, delayed update is the case  $(k+1-d)~{\rm mod}~\tau = 0 $. 

\textbf{Matrix Representation.}
Define matrices $\boldsymbol{X}_k$, $\boldsymbol{G}_k\in\mathbb{R}^{d\times M}$ that concatenate all local models and gradients in $k$ iteration:

\begin{equation}\nonumber
\boldsymbol{X}_k = \left[
x_{k}^{1}, ... , x_{k}^{m}
\right] 
,~
\boldsymbol{G}_k= \left[
g \left(  x_{k}^{(1)} \right), ... , g \left(  x_{k}^{(m)} \right)
\right] 
\end{equation}

Then, the update rule is
\begin{equation}\label{equ:theorem1_1}
\boldsymbol{X}_{k+1} =  \left \{
\begin{array}{ll}
\xi\left( \boldsymbol{X}_{k}-\eta \boldsymbol{G}_{k} \right) +(1-\xi) \left( \boldsymbol{X}_{k-d}-\eta \boldsymbol{G}_{k-d} \right) \boldsymbol{J}, &(k+1-d)\mod \tau = 0 \\
\boldsymbol{X}_{k}-\eta \boldsymbol{G}_{k}, \quad & {\rm otherwise} \\
\end{array}
\right.
\end{equation}

\textbf{Update Rule for the Averaged Model.}
The update rule of DaSGD is given by
\begin{equation}\nonumber
\begin{aligned}
x_{k+1}^{(m)}  = & \left \{
\begin{array}{ll}
x_{k}^{(m)} - \eta g \left(  x_{k}^{(m)} \right)   , \quad & {\rm otherwise} \\
\xi x_{k}^{(m)} - \eta\xi g \left(  x_{k}^{(m)} \right) +\frac{1-\xi}{M}\sum\limits_{j=1}^{M}\left[  x_{k-d}^{(j)} - \eta g \left(  x_{k-d}^{(j)} \right)  \right]     , &(k+1-d)\mod \tau = 0 \\
\end{array}
\right.\\
\end{aligned}
\end{equation}

Here, we set
\begin{equation} \nonumber\bar{x}_k =\frac{1}{M} \sum_{i=1}^{M}   x_{k}^{(i)} ,~ \bar{g}_k = \frac{1}{M} \sum_{i=1}^{M} g \left(  x_{k}^{(i)} \right)\end{equation}

The average weight on different workers is obtained by
\begin{equation}\nonumber
\begin{aligned}
\bar{x}_{k+1} = &\left \{
\begin{array}{ll}
\bar{x}_{k} - \eta\bar{g}_k   , \quad & {\rm otherwise} \\
\xi \bar{x}_{k} + (1-\xi)\bar{x}_{k-d} - \eta\xi \bar{g}_k  -\eta(1-\xi)\bar{g}(x_{k-d}), &(k+1-d)\mod \tau = 0 \\
\end{array}
\right.\\
\end{aligned}
\end{equation}

When $ z = \tau(k+1)$ for $z\mod \tau = 0$, we have
\begin{equation}\nonumber
\begin{aligned}
\bar{x}_{\tau(k+1)+d} =~&  \xi\bar{x}_{\tau(k+1)+d-1} + (1-\xi)\bar{x}_{\tau(k+1)-1} - \xi\eta\bar{g}_{\tau(k+1)+d-1}-(1-\xi)\eta \bar{g}_{\tau(k+1)-1}\\
=~&  \xi\bar{x}_{\tau k+d} +(1-\xi)\bar{x}_{\tau k+d}
-\xi\eta \sum\limits_{i=0}^{\tau-1} \bar{g}_{\tau k+d+i}
- (1-\xi)\eta \sum\limits_{i=0}^{\tau-1-d}  \bar{g}_{\tau k+d+i} \\
=~& \bar{x}_{\tau k+d}
-\eta \left[ \xi  \left( \sum\limits_{i=0}^{\tau-1} \bar{g}_{\tau k+d+i}-\sum\limits_{i=0}^{\tau-1-d}  \bar{g}_{\tau k+d+i} \right) 
+ \sum\limits_{i=0}^{\tau-1-d}  \bar{g}_{\tau k+d+i} \right] \\
=~& \bar{x}_{\tau k+d}
-\eta \left[ \xi   \sum\limits_{i=\tau-d}^{\tau-1} \bar{g}_{\tau k+d+i}
+ \sum\limits_{i=0}^{\tau-1-d}  \bar{g}_{\tau k+d+i} \right] \\
\end{aligned}
\end{equation}

If we set $K(k) = \tau k+d$
\begin{equation}\nonumber
\begin{aligned}
\bar{x}_{K(k+1)} =~& \bar{x}_{K(k)}
-\eta \left[ 
\xi   \sum\limits_{i=\tau-d}^{\tau-1} \bar{g}_{K(k)+i}
+ \sum\limits_{i=0}^{\tau-1-d}  \bar{g}_{K(k)+i}
\right] \\
\end{aligned}
\end{equation}

For the ease of writing, we first define some notations. Let $\mathcal{S}_k$ denote the set $\left\lbrace s_k^{(1)},...,s_k^{(m)} \right\rbrace $ of mini-batches at $m$ workers in iteration $k$. Besides, define averaged stochastic gradient and averaged full batch gradient as follows:

\begin{equation}\label{equ:G_kappa}
\mathcal{G}_{K(k)}  =
\frac{1}{M}\sum_{m=1}^{M}
\left[ \sum\limits_{i=\tau-d}^{\tau-1}   \xi  g \left(  x_{\tau k+d+i}^{(m)} \right)
+\sum\limits_{i=0}^{\tau-1-d} g \left(  x_{\tau k+d+i}^{(m)}\right)\right] \end{equation}
\begin{equation}\label{equ:H_kappa}
\mathcal{H}_{K(k)}  =
\frac{1}{M} \sum_{m=1}^{M} \left[ 
\sum\limits_{i=\tau-d}^{\tau-1} \xi \bigtriangledown F \left(  x_{\tau k+d+i}^{(m)} \right)
+\sum\limits_{i=0}^{\tau-1-d}  \bigtriangledown F \left(  x_{\tau k+d+i}^{(m)}\right)\right] \end{equation}
\begin{equation}\label{equ:X_kappa} 
\mu_{K(k)} =\frac{1}{M} \sum_{i=1}^{M}   x_{\tau k+d}^{(i)}\end{equation}

Then we have
\begin{equation}\nonumber
\mu_{K(k+1)} =~ \mu_{K(k)}-\eta \mathcal{G}_{K(k)}
\end{equation}

\subsection{Convergence Rate}

\textbf{Theorem (Convergence of DaSGD).} 
Under assumptions, if the learning rate satisfies 
\begin{equation}\nonumber
\eta  \leq \min \left\lbrace \sqrt{ a} , \sqrt{ b} \right\rbrace 
\end{equation}
where $a=1/ \left\lbrace 2L\xi^2(\beta+1)(1-\xi)+  6 L^2 (d \xi+\tau -d)[ (\beta+k\tau)+(\beta+1)(1-\xi) ] \right\rbrace $, $b=\xi M(1-\xi)/\left\lbrace 2L\xi^2(\beta+1)(1-\xi)+ 3L^2 M(\tau-d) (2\beta+2k\tau)    +  6dM \xi L^2 [ (\beta+k\tau)+(\beta+1)(1-\xi) ]  \right\rbrace $.
Then the average-squared gradient norm after $K$ iterations is bounded as
\begin{equation}\begin{aligned}\nonumber
&\mathbb{E}\left[ \frac{1}{K}\sum\limits_{k=1}^{K}  
\left\|  \bigtriangledown F(\mu_k)\right\|^2 \right] \\
\leq &\frac{2M\left[ F(\mu_1) - F_{inf}\right]   
	+   2MKL\eta^2\sigma^2 \left[ \xi^2 d + \tau-d \right]   }{\eta MK(\xi d+\tau-d)}  
+\frac{3\eta^4 \xi L^2  (\tau-d+d\xi )   }{MK(\xi d+\tau-d )} \frac{\xi^{2}}{1-\xi^{2}}   \left\| \sum\limits_{i=1}^{d-1} g(\textbf{X}_{d-1}) \right\|^2_F  
\\
&+ \frac{6 \eta^4 L^2 \sigma^2  }{\xi d+\tau-d  }   \left( 
\tau\frac{\xi^{2}}{1-\xi^{2}}(\tau-d+\xi d)
+(\tau-d)^2+\xi d(\tau -1)
\right)
\end{aligned}\end{equation}
where $\mu_k=\frac{1}{M} \sum_{i=1}^{M}   x_{\tau k+d}^{(i)}$, $\left\|~\right\|_F^2 $ is the Frobenius norm.

\textbf{Corollary.} 
Under sssumptions, if the learning rate is $\eta= A/\sqrt{K}$ the average-squared gradient norm after $K$ iterations is bounded by
\begin{equation}\begin{aligned}\nonumber
&\mathbb{E}\left[ \frac{1}{K}\sum\limits_{k=1}^{K}  
\left\|  \bigtriangledown F(\mu_k)\right\|^2 \right] \\
\leq &\frac{2M\left[ F(\mu_1) - F_{inf}\right]   
	+   2MLA^2\sigma^2 \left[ \xi^2 d + \tau-d \right]   }{AM\sqrt{K}(\xi d+\tau-d)}  
+\frac{3A^4 \xi L^2  (\tau-d+d\xi )   }{MK^3(\xi d+\tau-d )} \frac{\xi^{2}}{1-\xi^{2}}   \left\| \sum\limits_{i=1}^{d-1} g(\textbf{X}_{d-1}) \right\|^2_F  
\\
&+ \frac{6 A^4 L^2 \sigma^2  }{K^2(\xi d+\tau-d)  }   \left( 
\tau\frac{\xi^{2}}{1-\xi^{2}}(\tau-d+\xi d)
+(\tau-d)^2+\xi d(\tau -1)
\right).
\end{aligned}\end{equation}

If the total iterations $K$ is sufficiently large, then the average-squared gradient norm is bounded by 
$$\mathbb{E}\left[ \frac{1}{K}\sum\limits_{k=1}^{K}  
\left\|  \bigtriangledown F(\mu_k)\right\|^2 \right] 
\leq
\frac{2M\left[ F(\mu_1) - F_{inf}\right]   
	+   2MLA^2\sigma^2 \left[ \xi^2 d + \tau-d \right]   }{AM\sqrt{K}(\xi d+\tau-d)}. $$

\subsection{Proof of Convergence Rate}
\textbf{Lemma 1.}
If the learning rate satisfies $\eta\leq M/[2L\xi^2(\beta+1)]$ and all local model parameters are initialized at the same point, then the average-squared gradient after $K$ iterations is bounded as follows
\begin{equation}\begin{aligned}\nonumber
&\mathbb{E}_{K(k)}\left[ \frac{1}{K}\sum\limits_{k=1}^{K}  
\left\|  \bigtriangledown F(\mu_{K(k)})\right\|^2 \right] \\
\leq &\frac{2\left[ F(\mu_1) - F_{inf}\right] }{\eta K(\xi d+\tau-d)} 
+ \frac{2L\eta\sigma^2 \left[ \xi^2 d + \tau-d \right]}{M(\xi d+\tau-d)} \\
&  + \frac{\eta^2 L^2 }{KM(\xi d+\tau-d ) }  \sum\limits_{k=1}^{K}  \sum_{m=1}^{M}   \left[      
\sum\limits_{i=0}^{\tau-1-d} \mathbb{E}_{K(k)} \left\|  \mu_{K(k)}- \ x_{\tau k+d+i}^{(m)} \right\|^2 
+\xi \sum\limits_{i=\tau-d}^{\tau-1}  
\mathbb{E}_{K(k)}\left\| \mu_{K(k)}- \ x_{\tau k+d+i}^{(m)} \right\|^2 \right] 
\end{aligned}\end{equation}

Proof.

From the Lipschitzisan gradient assumption $ ||\bigtriangledown F(x)- \bigtriangledown F(y) || \leq  L||x-y||$, we have

\begin{equation}\label{equ:lipschitzisan_gradient2}
\begin{aligned}
F(X_{K(k+1)})-F(X_{K(k)})  \leq& \left\langle  \bigtriangledown F(X_{K(k)}), X_{K(k+1)}-X_{K(k)}\right\rangle + \frac{L}{2} \left\|X_{K(k+1)}-X_{K(k)}\right\|^2 \\
= &-\eta  \left\langle  \bigtriangledown F(X_{K(k)}), \mathcal{G}_{K(k)}\right\rangle + \frac{L\eta^2}{2} \left\| \mathcal{G}_{K(k)}\right\|^2 
\end{aligned}
\end{equation}

Taking expectation respect to $\mathcal{S}_{K(k)}$ on both sides of \eqref{equ:lipschitzisan_gradient2}, we have
\begin{equation}\nonumber
\begin{aligned}\nonumber
\mathbb{E}_{K(k)}\left[ F(X_{K(k+1)})\right] -F(X_{K(k)})  \leq 
& -\eta \mathbb{E}_{K(k)} \left[ \left\langle  \bigtriangledown F(X_{K(k)}), \mathcal{G}_{K(k)}\right\rangle \right]+
\frac{L\eta^2}{2}\mathbb{E}_{K(k)} \left[ \left\| \mathcal{G}_{K(k)}\right\|^2 \right] 
\end{aligned}
\end{equation}

From the fact
\begin{equation}\nonumber
\left\langle a,b\right\rangle = \frac{1}{2}\left( ||a||^2+||b||^2 -||a-b||^2 \right) 
\end{equation}

we have
\begin{align}\nonumber
\mathbb{E}_{K(k)}\left[ F(X_{K(k+1)})\right] -F(X_{K(k)})  \leq 
& -\eta \mathbb{E}_{K(k)} \left[ \left\langle  \bigtriangledown F(X_{K(k)}), \mathcal{G}_{K(k)}\right\rangle \right]+
\frac{L\eta^2}{2}\mathbb{E}_{K(k)} \left[ \left\| \mathcal{G}_{K(k)}\right\|^2 \right] 
\end{align}

Combining with Lemmas 4 and 5, we obtain
\begin{align}
&\mathbb{E}_{K(k)}\left[ F(X_{K(k+1)})\right] -F(X_{K(k)})  \\
\leq & -\eta \mathbb{E}_{K(k)} \left[ \left\langle  \bigtriangledown F(X_{K(k)}), \mathcal{G}_{K(k)}\right\rangle \right]+
\frac{L\eta^2}{2}\mathbb{E}_{K(k)} \left[ \left\| \mathcal{G}_{K(k)}\right\|^2 \right] \\
\leq &- \eta \frac{\xi d+\tau-d}{2} \left\|  \bigtriangledown F(X_{K(k)})\right\|^2 \\
&+\left[ \frac{L\xi^2\eta^2(\beta+1)}{M^2}- \frac{\eta\xi }{2M} \right] 
\sum\limits_{i=\tau-d}^{\tau-1}  \left\| \bigtriangledown F\left(  \textbf{X}_{\tau k+d+i}\right) \right\|^2_F
+\left[ \frac{L\xi^2\eta^2(\beta+1)}{M^2}- \frac{\eta}{2M} \right] 
\sum\limits_{i=0}^{\tau-1-d}  \left\| \bigtriangledown F\left(  \textbf{X}_{\tau k+d+i}\right) \right\|^2_F \\
&  +\eta \frac{1 }{2M} \sum_{m=1}^{M}\left[ \sum\limits_{i=0}^{\tau-1-d}  \left\| \bigtriangledown F(X_{K(k)})- \bigtriangledown F \left(  x_{\tau k+d+i}^{(m)} \right)\right\|^2 +\xi\sum\limits_{i=\tau-d}^{\tau-1}  \left\| \bigtriangledown F(X_{K(k)})- \bigtriangledown F \left(  x_{\tau k+d+i}^{(m)} \right)\right\|^2 
\right] \\
&+ \frac{L\eta^2\sigma^2 \left[ \xi^2 d + \tau-d \right]}{M} \\
\leq &- \eta \frac{\xi d+\tau-d}{2} \left\|  \bigtriangledown F(X_{K(k)})\right\|^2 \\
&+\left[ \frac{L\xi^2\eta^2(\beta+1)}{M^2}- \frac{\eta\xi }{2M} \right] 
\sum\limits_{i=\tau-d}^{\tau-1}  \left\| \bigtriangledown F\left(  \textbf{X}_{\tau k+d+i}\right) \right\|^2_F
+\left[ \frac{L\xi^2\eta^2(\beta+1)}{M^2}- \frac{\eta}{2M} \right] 
\sum\limits_{i=0}^{\tau-1-d}  \left\| \bigtriangledown F\left(  \textbf{X}_{\tau k+d+i}\right) \right\|^2_F \\
&  + \frac{\eta L^2 }{2M} \sum\limits_{i=0}^{\tau-1-d}  \sum_{m=1}^{M}\left\| 
\mu_{K(k)}- \ x_{\tau k+d+i}^{(m)} \right\|^2 
+\frac{\eta \xi L^2}{2M} \sum\limits_{i=\tau-d}^{\tau-1}  \sum_{m=1}^{M}\left\| 
\mu_{K(k)}- \ x_{\tau k+d+i}^{(m)} \right\|^2  \label{equ:lemma3_3}\\
&+ \frac{L\eta^2\sigma^2 \left[ \xi^2 d + \tau-d \right]}{M} 
\end{align}

where \eqref{equ:lemma3_3} is due to the Lipschitzisan gradient assumption $ ||\bigtriangledown F(x)- \bigtriangledown F(y) || \leq  L||x-y||$.
After minor rearranging and according to the definition of Frobenius norm, it is easy to show
\begin{align}\nonumber
& \eta \frac{\xi d+\tau-d}{2} \left\|  \bigtriangledown F(\mu_{K(k)})\right\|^2 \\
\leq &F(\mu_{K(k)}) -\mathbb{E}_{K(k)}\left[ F(\mu_{K(k+1)})\right]  + \frac{L\eta^2\sigma^2 \left[ \xi^2 d + \tau-d \right]}{M} \\
&+\left[ \frac{L\xi^2\eta^2(\beta+1)}{M^2}- \frac{\eta\xi }{2M} \right] 
\sum\limits_{i=\tau-d}^{\tau-1}  \left\| \bigtriangledown F\left(  \textbf{X}_{\tau k+d+i}\right) \right\|^2_F
+\left[ \frac{L\xi^2\eta^2(\beta+1)}{M^2}- \frac{\eta}{2M} \right] 
\sum\limits_{i=0}^{\tau-1-d}  \left\| \bigtriangledown F\left(  \textbf{X}_{\tau k+d+i}\right) \right\|^2_F \\
&  + \frac{\eta L^2 }{2M} \sum\limits_{i=0}^{\tau-1-d}  \sum_{m=1}^{M}\left\| 
\mu_{K(k)}- \ x_{\tau k+d+i}^{(m)} \right\|^2 
+\frac{\eta \xi L^2}{2M} \sum\limits_{i=\tau-d}^{\tau-1}  \sum_{m=1}^{M}\left\| 
\mu_{K(k)}- \ x_{\tau k+d+i}^{(m)} \right\|^2 
\end{align}

Taking the total expectation and averaging over all iterates, we have
\begin{equation}\begin{aligned}\nonumber
& \eta \frac{\xi d+\tau-d}{2} \mathbb{E}_{K(k)}\left[ \frac{1}{K}\sum\limits_{k=1}^{K}  
\left\|  \bigtriangledown F(\mu_{K(k)})\right\|^2 \right] \\
\leq &\frac{F(\mu_1) - F_{inf}}{K} 
+ \frac{L\eta^2\sigma^2 \left[ \xi^2 d + \tau-d \right]}{M} \\
&+\left[ \frac{L\xi^2\eta^2(\beta+1)}{KM^2}- \frac{\eta\xi }{2KM} \right] 
\sum\limits_{k=1}^{K} \sum\limits_{i=\tau-d}^{\tau-1}  \mathbb{E}_{K(k)}\left\| \bigtriangledown F\left(  \textbf{X}_{\tau k+d+i}\right) \right\|^2_F
+\left[ \frac{L\xi^2\eta^2(\beta+1)}{KM^2}- \frac{\eta}{2KM} \right] 
\sum\limits_{k=1}^{K} \sum\limits_{i=0}^{\tau-1-d}  \mathbb{E}_{K(k)} \left\| \bigtriangledown F\left(  \textbf{X}_{\tau k+d+i}\right) \right\|^2_F \\
&  + \frac{\eta L^2 }{2KM} \sum\limits_{k=1}^{K} \sum\limits_{i=0}^{\tau-1-d}  \sum_{m=1}^{M}        
\mathbb{E}_{K(k)} \left\|  \mu_{K(k)}- \ x_{\tau k+d+i}^{(m)} \right\|^2 
+\frac{\eta \xi L^2}{2KM} \sum\limits_{k=1}^{K} \sum\limits_{i=\tau-d}^{\tau-1}  \sum_{m=1}^{M}
\mathbb{E}_{K(k)}\left\| \mu_{K(k)}- \ x_{\tau k+d+i}^{(m)} \right\|^2 
\end{aligned}\end{equation}

Then, we have
\begin{equation}\begin{aligned}
\mathbb{E}_{K(k)}\left[ \frac{1}{K}\sum\limits_{k=1}^{K}  
\left\|  \bigtriangledown F(\mu_{K(k)})\right\|^2 \right] 
\leq &\frac{2\left[ F(\mu_1) - F_{inf}\right] }{\eta K(\xi d+\tau-d)} 
+ \frac{2L\eta\sigma^2 \left[ \xi^2 d + \tau-d \right]}{M(\xi d+\tau-d)} \\
&+ \frac{2L\xi^2\eta^4(\beta+1)-\eta^2\xi M}{KM^2 (\xi d+\tau-d )}  
\sum\limits_{k=1}^{K} \sum\limits_{i=\tau-d}^{\tau-1}  \mathbb{E}_{K(k)}\left\| \bigtriangledown F\left(  \textbf{X}_{\tau k+d+i}\right) \right\|^2_F\\
&+ \frac{2L\xi^2\eta^4(\beta+1)-\eta^2 M}{KM^2 (\xi d+\tau-d )} 
\sum\limits_{k=1}^{K} \sum\limits_{i=0}^{\tau-1-d}  \mathbb{E}_{K(k)} \left\| \bigtriangledown F\left(  \textbf{X}_{\tau k+d+i}\right) \right\|^2_F \\
&  + \frac{\eta^2 L^2 }{KM(\xi d+\tau-d ) }  \sum\limits_{k=1}^{K} \sum\limits_{i=0}^{\tau-1-d}  \sum_{m=1}^{M}        
\mathbb{E}_{K(k)} \left\|  \mu_{K(k)}- \ x_{\tau k+d+i}^{(m)} \right\|^2 \\
&+\frac{\eta^2 \xi L^2}{KM(\xi d+\tau-d )} \sum\limits_{k=1}^{K} \sum\limits_{i=\tau-d}^{\tau-1}  \sum_{m=1}^{M}
\mathbb{E}_{K(k)}\left\| \mu_{K(k)}- \ x_{\tau k+d+i}^{(m)} \right\|^2 \label{equ:lemma3_4}
\end{aligned}\end{equation}

If the learning rate satisfies $\eta\leq \sqrt{\frac{M}{2L\xi^2(\beta+1)}}$, then
\begin{equation}\begin{aligned}\nonumber
\mathbb{E}_{K(k)}\left[ \frac{1}{K}\sum\limits_{k=1}^{K}  
\left\|  \bigtriangledown F(\mu_{K(k)})\right\|^2 \right] 
\leq &\frac{2\left[ F(\mu_1) - F_{inf}\right] }{\eta K(\xi d+\tau-d)} 
+ \frac{2L\eta\sigma^2 \left[ \xi^2 d + \tau-d \right]}{M(\xi d+\tau-d)} \\
&  + \frac{\eta^2 L^2 }{KM(\xi d+\tau-d ) }  \sum\limits_{k=1}^{K} \sum\limits_{i=0}^{\tau-1-d}  \sum_{m=1}^{M}        
\mathbb{E}_{K(k)} \left\|  \mu_{K(k)}- \ x_{\tau k+d+i}^{(m)} \right\|^2 \\
&+\frac{\eta^2 \xi L^2}{KM(\xi d+\tau-d )} \sum\limits_{k=1}^{K} \sum\limits_{i=\tau-d}^{\tau-1}  \sum_{m=1}^{M}
\mathbb{E}_{K(k)}\left\| \mu_{K(k)}- \ x_{\tau k+d+i}^{(m)} \right\|^2 
\end{aligned}\end{equation}

Recalling the definition $\mu_{K(k)}=\frac{1}{M} \sum_{i=1}^{M}   x_{\tau k+d}^{(i)} = \textbf{X}_{K(k)} \textbf{1}_M/M$ and adding a positive term to the RHS, one can get
\begin{align}\nonumber
\sum\limits_{i=\tau-d}^{\tau-1} \sum_{m=1}^{M}\left\| \mu_{K(k)}- \ x_{\tau k+d+i}^{(m)} \right\|^2 =& \sum\limits_{i=\tau-d}^{\tau-1}\left\| \textbf{X}_{\tau k+d}\textbf{J}-\textbf{X}_{\tau k+d+i}\right\|^2_F
\end{align}

We have
\begin{equation}\begin{aligned}\nonumber
\mathbb{E}_{K(k)}\left[ \frac{1}{K}\sum\limits_{k=1}^{K}  
\left\|  \bigtriangledown F(\mu_{K(k)})\right\|^2 \right] 
\leq &\frac{2\left[ F(\mu_1) - F_{inf}\right] }{\eta K(\xi d+\tau-d)} 
+ \frac{2L\eta\sigma^2 \left[ \xi^2 d + \tau-d \right]}{M(\xi d+\tau-d)} \\
&  + \frac{\eta^2 L^2 }{KM(\xi d+\tau-d ) }  \sum\limits_{k=1}^{K} \sum\limits_{i=0}^{\tau-1-d}    
\mathbb{E}_{K(k)} \left\| \textbf{X}_{\tau k+d}\textbf{J}-\textbf{X}_{\tau k+d+i}\right\|^2_F\\
&+\frac{\eta^2 \xi L^2}{KM(\xi d+\tau-d )} \sum\limits_{k=1}^{K} \sum\limits_{i=\tau-d}^{\tau-1} 
\mathbb{E}_{K(k)}\left\| \textbf{X}_{\tau k+d}\textbf{J}-\textbf{X}_{\tau k+d+i}\right\|^2_F
\end{aligned}\end{equation}

\bigskip \hrule  \bigskip

\textbf{Lemma 2}.

\begin{equation}\left\| \mathcal{H}_{K(k)}\right\|^2\leq
\frac{2d\xi^2}{M}   \sum\limits_{i=\tau-d}^{\tau-1}     \left\| \bigtriangledown F\left(  \textbf{X}_{\tau k+d+i}\right) \right\|^2_F
+\frac{2(\tau-d)}{M} \sum\limits_{i=0}^{\tau-1-d}   \left\| \bigtriangledown F\left(  \textbf{X}_{\tau k+d+i}\right) \right\|^2_F   \label{equ:lemma3}
\end{equation}

Proof. 
\begin{align}
\left\| \mathcal{H}_{K(k)}\right\|^2 =~&\left\|  
\xi  \frac{1}{M}\sum\limits_{i=\tau-d}^{\tau-1}  \sum_{m=1}^{M} \bigtriangledown F \left(  x_{\tau k+d+i}^{(m)} \right)
+\frac{1}{M}\sum\limits_{i=0}^{\tau-1-d}  \sum_{m=1}^{M}  \bigtriangledown F \left(  x_{\tau k+d+i}^{(m)}\right)
\right\|^2  \\ 
\leq~& \frac{2d\xi^2}{M^2}  
\sum\limits_{i=\tau-d}^{\tau-1} \left\| \sum_{m=1}^{M} \bigtriangledown F \left(  x_{\tau k+d+i}^{(m)} \right)\right\|^2
+\frac{2(\tau-d)}{M^2} \sum\limits_{i=0}^{\tau-1-d} \left\|  \sum_{m=1}^{M}  \bigtriangledown F \left(  x_{\tau k+d+i}^{(m)}\right)
\right\|^2 \label{equ:lemma3_1} \\ 
\leq~& \frac{2d\xi^2}{M}  
\sum\limits_{i=\tau-d}^{\tau-1} \sum_{m=1}^{M}\left\|  \bigtriangledown F \left(  x_{\tau k+d+i}^{(m)} \right)\right\|^2
+\frac{2(\tau-d)}{M} \sum\limits_{i=0}^{\tau-1-d}  \sum_{m=1}^{M}  \left\|  \bigtriangledown F \left(  x_{\tau k+d+i}^{(m)}\right)
\right\|^2 \label{equ:lemma3_2} \\ 
=~&  \frac{2d\xi^2}{M}   \sum\limits_{i=\tau-d}^{\tau-1}     \left\| \bigtriangledown F\left(  \textbf{X}_{\tau k+d+i}\right) \right\|^2_F
+\frac{2(\tau-d)}{M} \sum\limits_{i=0}^{\tau-1-d}   \left\| \bigtriangledown F\left(  \textbf{X}_{\tau k+d+i}\right) \right\|^2_F
\end{align}

where \eqref{equ:lemma3_1} is due to $\left\| a+b \right\|^2 \leq 2 \left\| a\right\|^2+2\left\| b \right\|^2$, \eqref{equ:lemma3_2} comes from the convexity of vector norm and Jensen’s inequality.

\bigskip \hrule  \bigskip

\textbf{Lemma 3}. Under assumptions $ \mathbb{E}_{\mathcal{S}_k| x}\left[ g(x)\right]=   \bigtriangledown F(x)$ and $  \mathbb{E}_{\mathcal{S}_k | x} ||g(x)-\bigtriangledown F(x)  ||^2 \leq \beta  ||\bigtriangledown F(x)||^2+\sigma^2  $, we have the following variance bound for the averaged stochastic gradient:
\begin{equation}
\mathbb{E}_{{K(k)}}\left[\left\|  \mathcal{G}_{K(k)} -  \mathcal{H}_{K(k)} \right\|^2 \right] \leq \frac{2\sigma^2 \left[ \xi^2 d + \tau-d \right]}{M} +  \frac{2\beta\xi^2}{M^2}   \sum\limits_{i=\tau-d}^{\tau-1}     \left\| \bigtriangledown F\left(  \textbf{X}_{\tau k+d+i}\right) \right\|^2_F
+\frac{2\beta}{M^2} \sum\limits_{i=0}^{\tau-1-d}   \left\| \bigtriangledown F\left(  \textbf{X}_{\tau k+d+i}\right) \right\|^2_F \label{equ:lemma4}
\end{equation}

Proof. According to the definition of \eqref{equ:G_kappa}, \eqref{equ:H_kappa}, and \eqref{equ:X_kappa}, we have
\begin{align}
&\mathbb{E}_{{K(k)}}\left[\left\|  \mathcal{G}_{K(k)} -  \mathcal{H}_{K(k)} \right\|^2 \right] \\ 
=~& \frac{1}{M^2} \mathbb{E}_{{K(k)}}\left[\left\| 
\xi  \sum\limits_{i=\tau-d}^{\tau-1}  \sum_{m=1}^{M} \left[  g \left(  x_{\tau k+d+i}^{(m)} \right)- \bigtriangledown F \left(  x_{\tau k+d+i}^{(m)} \right)\right] 
+\sum\limits_{i=0}^{\tau-1-d}  \sum_{m=1}^{M} \left[  g \left(  x_{\tau k+d+i}^{(m)}\right) -  \bigtriangledown F \left(  x_{\tau k+d+i}^{(m)}\right)\right] 
\right\|^2 \right] \\
\leq~& \frac{2}{M^2} \mathbb{E}_{K(k)}\left[\left\| 
\xi  \sum\limits_{i=\tau-d}^{\tau-1}  \sum_{m=1}^{M} \left[  g \left(  x_{\tau k+d+i}^{(m)} \right)- \bigtriangledown F \left(  x_{\tau k+d+i}^{(m)} \right)\right] \right\|^2
+ \left\| \sum\limits_{i=0}^{\tau-1-d}  \sum_{m=1}^{M} \left[  g \left(  x_{\tau k+d+i}^{(m)}\right) -  \bigtriangledown F \left(  x_{\tau k+d+i}^{(m)}\right)\right] 
\right\|^2 \right] \label{equ:lemma4_1} \\ 
=~& \frac{2}{M^2} \mathbb{E}_{K(k)}\left[
\xi^2  \sum\limits_{i=\tau-d}^{\tau-1}  \sum_{m=1}^{M} \left\|  g \left(  x_{\tau k+d+i}^{(m)} \right)- \bigtriangledown F \left(  x_{\tau k+d+i}^{(m)} \right) \right\|^2 
+ \sum\limits_{i=0}^{\tau-1-d}  \sum_{m=1}^{M} \left\|  g \left(  x_{\tau k+d+i}^{(m)}\right) -  \bigtriangledown F \left(  x_{\tau k+d+i}^{(m)}\right)
\right\|^2
\right. \\
&+ \xi^2  \sum\limits_{j\neq i}^{\tau-1}  \sum_{l\neq m}^{M} \left\langle 
g \left(  x_{\tau k+d+i}^{(m)} \right)- \bigtriangledown F \left(  x_{\tau k+d+i}^{(m)} \right), g \left(  x_{\tau k+d+j}^{(l)} \right)- \bigtriangledown F \left(  x_{\tau k+d+j}^{(l)} \right) 
\right\rangle \label{equ:lemma4_2} \\
&+\left. \sum\limits_{j\neq i}^{\tau-1-d}  \sum_{l\neq m}^{M} \left\langle 
g \left(  x_{\tau k+d+i}^{(m)} \right)- \bigtriangledown F \left(  x_{\tau k+d+i}^{(m)} \right), g \left(  x_{\tau k+d+j}^{(l)} \right)- \bigtriangledown F \left(  x_{\tau k+d+j}^{(l)} \right) 
\right\rangle 
\right]  \label{equ:lemma4_3}  \\
=~& \frac{2\xi^2}{M^2}   \sum\limits_{i=\tau-d}^{\tau-1}  \sum_{m=1}^{M} 
\mathbb{E}_{K(k)} \left\|  g \left(  x_{\tau k+d+i}^{(m)} \right)- \bigtriangledown F \left(  x_{\tau k+d+i}^{(m)} \right) \right\|^2 
+ \frac{2}{M^2}  \sum\limits_{i=0}^{\tau-1-d}  \sum_{m=1}^{M} \mathbb{E}_{K(k)} \left\|  g \left(  x_{\tau k+d+i}^{(m)}\right) -  \bigtriangledown F \left(  x_{\tau k+d+i}^{(m)}\right)
\right\|^2 \label{equ:lemma4_4}    
\end{align}

where \eqref{equ:lemma4_1} is due to $\left\| a+b \right\|^2 \leq 2 \left\| a\right\|^2+2\left\| b \right\|^2$, \eqref{equ:lemma4_4} is due to $s_k^{i}$ are independent random variables and the assumption $ \mathbb{E}_{\mathcal{S}_k | x}\left[ g(x)\right]=   \bigtriangledown F(x)$. Now, directly applying assumption $  \mathbb{E}_{\mathcal{S}_k| x} ||g(x)-\bigtriangledown F(x)  ||^2 \leq \beta  ||\bigtriangledown F(x)||^2+\sigma^2 $ to \eqref{equ:lemma4_4}. Then, we have
\begin{align}\nonumber
\mathbb{E}_{{K(k)}}\left[\left\|  \mathcal{G}_{K(k)} -  \mathcal{H}_{K(k)} \right\|^2 \right] 
\leq~& \frac{2\xi^2}{M^2}   \sum\limits_{i=\tau-d}^{\tau-1}  \sum_{m=1}^{M}  \left[ \beta  \left\| \bigtriangledown F( x_{\tau k+d+i}^{(m)})\right\|^2+\sigma^2\right] 
+\frac{2}{M^2} \sum\limits_{i=0}^{\tau-1-d}  \sum_{m=1}^{M}\left[ \beta  \left\| \bigtriangledown F( x_{\tau k+d+i}^{(m)})\right\|^2+\sigma^2\right]  \\
=~& \frac{2\sigma^2 \left[ \xi^2 d + \tau-d \right]}{M} +  \frac{2\xi^2}{M^2}   \sum\limits_{i=\tau-d}^{\tau-1}  \sum_{m=1}^{M}  \beta  \left\| \bigtriangledown F( x_{\tau k+d+i}^{(m)})\right\|^2 
+\frac{2}{M^2} \sum\limits_{i=0}^{\tau-1-d}  \sum_{m=1}^{M}\beta  \left\| \bigtriangledown F( x_{\tau k+d+i}^{(m)})\right\|^2 \\
=~& \frac{2\sigma^2 \left[ \xi^2 d + \tau-d \right]}{M} +  \frac{2\beta\xi^2}{M^2}   \sum\limits_{i=\tau-d}^{\tau-1}     \left\| \bigtriangledown F\left(  \textbf{X}_{\tau k+d+i}\right) \right\|^2_F
+\frac{2\beta}{M^2} \sum\limits_{i=0}^{\tau-1-d}   \left\| \bigtriangledown F\left(  \textbf{X}_{\tau k+d+i}\right) \right\|^2_F
\end{align}

\bigskip \hrule  \bigskip

\textbf{Lemma 4}. Under assumption $ \mathbb{E}_{\mathcal{S}_k | x}\left[ g(x)\right]=   \bigtriangledown F(x)$, the expected inner product between stochastic gradient and full batch gradient can be expanded as

\begin{equation}\nonumber
\begin{aligned}
&\mathbb{E}_{K(k)} \left[ \left\langle  \bigtriangledown F(X_{K(k)}), \mathcal{G}_{K(k)}\right\rangle \right] \\
=~&  \frac{\xi d+\tau-d}{2} \left\|  \bigtriangledown F(X_{K(k)})\right\|^2 
+\frac{1 }{2M}\left[ \xi\sum\limits_{i=\tau-d}^{\tau-1}  \left\| \bigtriangledown F\left(  \textbf{X}_{\tau k+d+i}\right) \right\|^2_F+\sum\limits_{i=0}^{\tau-1-d}  \left\| \bigtriangledown F\left(  \textbf{X}_{\tau k+d+i}\right) \right\|^2_F 
\right] \\
&  - \frac{1 }{2M} \sum_{m=1}^{M}  \left[ 
\sum\limits_{i=0}^{\tau-1-d}  \left\| \bigtriangledown F(X_{K(k)})- \bigtriangledown F \left(  x_{\tau k+d+i}^{(m)} \right)\right\|^2+\xi \sum\limits_{i=\tau-d}^{\tau-1}  \left\| \bigtriangledown F(X_{K(k)})- \bigtriangledown F \left(  x_{\tau k+d+i}^{(m)} \right)\right\|^2 
\right]  
\end{aligned}
\end{equation}

Proof.
\begin{align}
&\mathbb{E}_{K(k)} \left[ \left\langle  \bigtriangledown F(X_{K(k)}), \mathcal{G}_{K(k)}\right\rangle \right]\\
=~& \mathbb{E}_{K(k)} \left[\left\langle  \bigtriangledown F(X_{K(k)}), \xi  \frac{1}{M}\sum\limits_{i=\tau-d}^{\tau-1}  \sum_{m=1}^{M} g \left(  x_{\tau k+d+i}^{(m)} \right)
+\frac{1}{M}\sum\limits_{i=0}^{\tau-1-d}  \sum_{m=1}^{M}  g \left(  x_{\tau k+d+i}^{(m)}\right)
\right\rangle \right] \\
=~&\xi  \frac{1}{M}\sum\limits_{i=\tau-d}^{\tau-1}  \sum_{m=1}^{M}\left\langle  \bigtriangledown F(X_{K(k)}),   \bigtriangledown F \left(  x_{\tau k+d+i}^{(m)} \right)\right\rangle\
+\frac{1}{M}\sum\limits_{i=0}^{\tau-1-d}  \sum_{m=1}^{M} \left\langle \bigtriangledown F(X_{K(k)}),   \bigtriangledown F \left(  x_{\tau k+d+i}^{(m)}\right)
\right\rangle\\
=~& \frac{\xi }{2M}\sum\limits_{i=\tau-d}^{\tau-1}  \sum_{m=1}^{M}
\left[ \left\|  \bigtriangledown F(X_{K(k)})\right\|^2 + \left\|   \bigtriangledown F \left(  x_{\tau k+d+i}^{(m)} \right)\right\|^2 - \left\| \bigtriangledown F(X_{K(k)})- \bigtriangledown F \left(  x_{\tau k+d+i}^{(m)} \right)\right\|^2
\right] \label{equ:lemma5_1}\\
&+\frac{1}{2M}\sum\limits_{i=0}^{\tau-1-d}  \sum_{m=1}^{M} 
\left[ \left\|  \bigtriangledown F(X_{K(k)})\right\|^2 + \left\|   \bigtriangledown F \left(  x_{\tau k+d+i}^{(m)} \right)\right\|^2 - \left\| \bigtriangledown F(X_{K(k)})- \bigtriangledown F \left(  x_{\tau k+d+i}^{(m)} \right)\right\|^2\right]  \label{equ:lemma5_2}\\
=~& \frac{\xi d+\tau-d}{2} \left\|  \bigtriangledown F(X_{K(k)})\right\|^2 
+\frac{1 }{2M}\left[ \xi\sum\limits_{i=\tau-d}^{\tau-1}  \left\| \bigtriangledown F\left(  \textbf{X}_{\tau k+d+i}\right) \right\|^2_F+\sum\limits_{i=0}^{\tau-1-d}  \left\| \bigtriangledown F\left(  \textbf{X}_{\tau k+d+i}\right) \right\|^2_F 
\right]\\
& - \frac{1 }{2M} \sum_{m=1}^{M}  \left[ 
\sum\limits_{i=0}^{\tau-1-d}  \left\| \bigtriangledown F(X_{K(k)})- \bigtriangledown F \left(  x_{\tau k+d+i}^{(m)} \right)\right\|^2+\xi \sum\limits_{i=\tau-d}^{\tau-1}  \left\| \bigtriangledown F(X_{K(k)})- \bigtriangledown F \left(  x_{\tau k+d+i}^{(m)} \right)\right\|^2 
\right] 
\end{align}

where \eqref{equ:lemma5_1} and \eqref{equ:lemma5_2} come from $\left\langle a,b\right\rangle = \frac{1}{2}\left( ||a||^2+||b||^2 -||a-b||^2 \right) $.

\bigskip \hrule  \bigskip

\textbf{Lemma 5.} Under assumptions $ E_{\xi | x}\left[ g(x)\right]=   \bigtriangledown F(x)$ and $  E_{\xi | x} ||g(x)-\bigtriangledown F(x)  ||^2 \leq \beta  ||\bigtriangledown F(x)||^2+\sigma^2  $, the squared norm of stochastic gradient can be bounded as
\begin{equation}\nonumber
\mathbb{E}_{{K(k)}} \left[ \left\| \mathcal{G}_{K(k)}\right\|^2 \right] \leq
\frac{2\sigma^2 \left[ \xi^2 d + \tau-d \right]}{M} +  \frac{2(\beta+1)}{M^2} \left[   \xi^2\sum\limits_{i=\tau-d}^{\tau-1}     \left\| \bigtriangledown F\left(  \textbf{X}_{\tau k+d+i}\right) \right\|^2_F
+ \sum\limits_{i=0}^{\tau-1-d}   \left\| \bigtriangledown F\left(  \textbf{X}_{\tau k+d+i}\right) \right\|^2_F\right] 
\end{equation}

Proof.
\begin{align}
\mathbb{E}_{{K(k)}} \left[ \left\| \mathcal{G}_{K(k)}\right\|^2 \right]
=~&  \mathbb{E}_{{K(k)}} \left[ \left\| \mathcal{G}_{K(k)} - \mathbb{E}_{{K(k)}}[\mathcal{G}_{K(k)}] \right\|^2 \right]+\left\| \mathbb{E}_{{K(k)}}[\mathcal{G}_{K(k)}]\right\|^2\\
=~&  \mathbb{E}_{{K(k)}} \left[ \left\| \mathcal{G}_{K(k)} - \mathcal{H}_{K(k)} \right\|^2 \right]+\left\| \mathcal{H}_{K(k)}\right\|^2\\
\leq~&  \frac{2\sigma^2 \left[ \xi^2 d + \tau-d \right]}{M} +  \frac{2d\xi^2M+2\beta\xi^2}{M^2}   \sum\limits_{i=\tau-d}^{\tau-1}     \left\| \bigtriangledown F\left(  \textbf{X}_{\tau k+d+i}\right) \right\|^2_F
+\frac{2(\tau-d)M+2\beta}{M^2} \sum\limits_{i=0}^{\tau-1-d}   \left\| \bigtriangledown F\left(  \textbf{X}_{\tau k+d+i}\right) \right\|^2_F \label{equ:t2_1}   \\
=~&  \frac{2\sigma^2 \left[ \xi^2 d + \tau-d \right]}{M} +  \frac{2(\beta+1)}{M^2} \left[   \xi^2\sum\limits_{i=\tau-d}^{\tau-1}     \left\| \bigtriangledown F\left(  \textbf{X}_{\tau k+d+i}\right) \right\|^2_F
+ \sum\limits_{i=0}^{\tau-1-d}   \left\| \bigtriangledown F\left(  \textbf{X}_{\tau k+d+i}\right) \right\|^2_F\right] 
\end{align}

where \eqref{equ:t2_1} follows \eqref{equ:lemma3} and \eqref{equ:lemma4}.

\bigskip \hrule  \bigskip

\textbf{Theorem 1 (Convergence of SGD).} 
Under assumptions, if the learning rate satisfies the following two formulas at the same time
\begin{equation}\begin{aligned}\nonumber
\eta  \leq & \sqrt{ \frac{1}{2L\xi^2(\beta+1)(1-\xi)+ 3L^2(\tau-d)[(1-\xi)(2\beta+2) +(2\beta+2k\tau)    ]+  3d \xi L^2 [ (2\beta+2k\tau)+(2\beta+2)(1-\xi) ] }}\\
= & \sqrt{ \frac{1}{2L\xi^2(\beta+1)(1-\xi)+  6 L^2 (d \xi+\tau -d)[ (\beta+k\tau)+(\beta+1)(1-\xi) ] }}
\end{aligned}\end{equation}

\begin{align}\nonumber
\eta  \leq & \sqrt{ \frac{\xi M(1-\xi)}{2L\xi^2(\beta+1)(1-\xi)+ 3L^2 M(\tau-d) (2\beta+2k\tau)    +  6dM \xi L^2 [ (\beta+k\tau)+(\beta+1)(1-\xi) ]  }}
\end{align}

Then the average-squared gradient norm after $K$ iterations is bounded as
\begin{equation}\begin{aligned}\nonumber
&\mathbb{E}_{K(k)}\left[ \frac{1}{K}\sum\limits_{k=1}^{K}  
\left\|  \bigtriangledown F(\mu_{K(k)})\right\|^2 \right] \\
\leq &\frac{2M\left[ F(\mu_1) - F_{inf}\right]   
	+   2MKL\eta^2\sigma^2 \left[ \xi^2 d + \tau-d \right]   }{\eta MK(\xi d+\tau-d)}  +\frac{3\eta^4 \xi L^2  (\tau-d+d\xi )   }{MK(\xi d+\tau-d )} \frac{\xi^{2}}{1-\xi^{2}}   \left\| \sum\limits_{i=1}^{d-1} g(\textbf{X}_{d-1}) \right\|^2_F  \\
&+ \frac{6 \eta^4 L^2 \sigma^2  }{\xi d+\tau-d  }   \left( 
\tau\frac{\xi^{2}}{1-\xi^{2}}(\tau-d+\xi d)
+(\tau-d)^2+\xi d(\tau -1)
\right)
\end{aligned}\end{equation}

Proof.

Recall the intermediate result \eqref{equ:lemma3_4} in the proof of Lemma 1:
\begin{equation}\begin{aligned}\nonumber
\mathbb{E}_{K(k)}\left[ \frac{1}{K}\sum\limits_{k=1}^{K}  
\left\|  \bigtriangledown F(\mu_{K(k)})\right\|^2 \right] 
\leq &\frac{2\left[ F(\mu_1) - F_{inf}\right] }{\eta K(\xi d+\tau-d)} 
+ \frac{2L\eta\sigma^2 \left[ \xi^2 d + \tau-d \right]}{M(\xi d+\tau-d)} \\
&+ \frac{2L\xi^2\eta^3=4(\beta+1)-\eta^2\xi M}{KM^2 (\xi d+\tau-d )}  
\sum\limits_{k=1}^{K} \sum\limits_{i=\tau-d}^{\tau-1}  \mathbb{E}_{K(k)}\left\| \bigtriangledown F\left(  \textbf{X}_{\tau k+d+i}\right) \right\|^2_F\\
&+ \frac{2L\xi^2\eta^4(\beta+1)-\eta^2 M}{KM^2 (\xi d+\tau-d )} 
\sum\limits_{k=1}^{K} \sum\limits_{i=0}^{\tau-1-d}  \mathbb{E}_{K(k)} \left\| \bigtriangledown F\left(  \textbf{X}_{\tau k+d+i}\right) \right\|^2_F \\
&  + \frac{\eta^2 L^2 }{KM(\xi d+\tau-d ) }  \sum\limits_{k=1}^{K} \sum\limits_{i=0}^{\tau-1-d}  \mathbb{E}_{K(k)} \left\| \textbf{X}_{\tau k+d}\textbf{J}-\textbf{X}_{\tau k+d+i}\right\|^2_F\\
&+\frac{\eta^2 \xi L^2}{KM(\xi d+\tau-d )} \sum\limits_{k=1}^{K} \sum\limits_{i=\tau-d}^{\tau-1} \mathbb{E}_{K(k)} \left\| \textbf{X}_{\tau k+d}\textbf{J}-\textbf{X}_{\tau k+d+i}\right\|^2_F
\end{aligned}\end{equation}

Our goal is to provide an upper bound for the network error term $\sum\limits_{k=1}^{K} \sum\limits_{i=\tau-d}^{\tau-1} \mathbb{E}_{K(k)} \left\| \textbf{X}_{\tau k+d}\textbf{J}-\textbf{X}_{\tau k+d+i}\right\|^2_F$.  First of all, let us derive a specific expression for  $\textbf{X}_{\tau k+d}\textbf{J}-\textbf{X}_{\tau k+d+i}$.

According to the update rule  \eqref{equ:theorem1_1}, one can observe that
\begin{align}
&\textbf{X}_{\tau k+d}\textbf{J}-\textbf{X}_{\tau k+d+i}\\
=~&\textbf{X}_{\tau k+d}(\textbf{J} - \textbf{I}  )+\eta\sum\limits_{j=0}^{i}\textbf{G}_{\tau k+d+j}\\
=~&\xi\left( \textbf{X}_{\tau k+d-1}-\eta \textbf{G}_{\tau k+d-1} \right) (\textbf{J} - \textbf{I})
+(1-\xi) \left( \textbf{X}_{\tau k}-\eta \textbf{G}_{\tau k} \right) \textbf{J}(\textbf{J} - \textbf{I})+\eta\sum\limits_{j=0}^{i}\textbf{G}_{\tau k+d+j}\\
=~&\xi \textbf{X}_{\tau (k-1)+d} (\textbf{J} - \textbf{I})
-\xi\eta \sum\limits_{i=0}^{\tau -1} \textbf{G}_{\tau (k-1)+d+i} (\textbf{J} - \textbf{I})
+\eta\sum\limits_{j=0}^{i}\textbf{G}_{\tau k+d+j}\\
=~&\xi^2 \textbf{X}_{\tau (k-2)+d} (\textbf{J} - \textbf{I})
-\eta\sum\limits_{j=1}^{2}\sum\limits_{i=0}^{\tau -1} \xi^j \textbf{G}_{\tau (k-j)+d+i} (\textbf{J} - \textbf{I})
+\eta\sum\limits_{j=0}^{i}\textbf{G}_{\tau k+d+j}\\
=~&\xi^k \textbf{X}_{d} (\textbf{J} - \textbf{I})
-\eta\sum\limits_{j=1}^{k}\sum\limits_{i=0}^{\tau -1} \xi^j \textbf{G}_{\tau (k-j)+d+i} (\textbf{J} - \textbf{I})
+\eta\sum\limits_{j=0}^{i}\textbf{G}_{\tau k+d+j}\\
=~&\xi^k \left( \textbf{X}_{d-1}-\eta \textbf{G}_{d-1} \right) (\textbf{J} - \textbf{I})
-\eta\sum\limits_{j=1}^{k}\sum\limits_{i=0}^{\tau -1} \xi^j \textbf{G}_{\tau (k-j)+d+i} (\textbf{J} - \textbf{I})
+\eta\sum\limits_{j=0}^{i}\textbf{G}_{\tau k+d+j}\\
=~&\xi^k  \textbf{X}_{1}(\textbf{J} - \textbf{I})
-\eta \xi^k  \sum\limits_{i=1}^{d-1} \textbf{G}_{i} (\textbf{J} - \textbf{I})
-\eta\sum\limits_{j=1}^{k}\sum\limits_{i=0}^{\tau -1} \xi^j \textbf{G}_{\tau (k-j)+d+i} (\textbf{J} - \textbf{I})
+\eta\sum\limits_{j=0}^{i}\textbf{G}_{\tau k+d+j}\\
=~&-\eta \xi^k  \sum\limits_{i=1}^{d-1} \textbf{G}_{i} (\textbf{J} - \textbf{I})
-\eta\sum\limits_{j=1}^{k}\sum\limits_{i=0}^{\tau -1} \xi^j \textbf{G}_{\tau (k-j)+d+i} (\textbf{J} - \textbf{I})
+\eta\sum\limits_{j=0}^{i}\textbf{G}_{\tau k+d+j} \label{equ:theorem1_2}
\end{align}

where \eqref{equ:theorem1_2} follows the fact that all workers start from the same point at the beginning of each local update period. Accordingly, we have
\begin{align}
&\sum\limits_{i=\tau-d}^{\tau-1} \mathbb{E}_{K(k)} \left\| \textbf{X}_{\tau k+d}\textbf{J}-\textbf{X}_{\tau k+d+i}\right\|^2_F\\
=&\sum\limits_{l=\tau-d}^{\tau-1} \mathbb{E}_{K(k)} \left\|
-\eta \xi^k  \sum\limits_{i=1}^{d-1} \textbf{G}_{i} (\textbf{J} - \textbf{I})
-\eta\sum\limits_{j=1}^{k}\sum\limits_{i=0}^{\tau -1} \xi^j \textbf{G}_{\tau (k-j)+d+i} (\textbf{J} - \textbf{I})
+\eta\sum\limits_{i=0}^{l}\textbf{G}_{\tau k+d+i}
\right\|^2_F\\
\leq& 3\eta^2  \mathbb{E}_{K(k)}\left[ 
\xi^{2k}d  \left\| \sum\limits_{i=1}^{d-1} \textbf{G}_{i} (\textbf{J} - \textbf{I})\right\|^2_F
+d\left\| \sum\limits_{j=1}^{k}\sum\limits_{i=0}^{\tau -1} \xi^j \textbf{G}_{\tau (k-j)+d+i} (\textbf{J} - \textbf{I})\right\|^2_F
+\sum\limits_{l=\tau-d}^{\tau-1} \left\| \sum\limits_{i=0}^{l}\textbf{G}_{\tau k+d+i}
\right\|^2_F\right] \\
\leq&3 \eta^2   \mathbb{E}_{K(k)}\left[ 
\xi^{2k}  d\left\| \sum\limits_{i=1}^{d-1} \textbf{G}_{d-1} \right\|^2_F
+d\left\| \sum\limits_{j=1}^{k}\sum\limits_{i=0}^{\tau -1} \xi^j \textbf{G}_{\tau (k-j)+d+i} \right\|^2_F
+\sum\limits_{l=\tau-d}^{\tau-1} \left\| \sum\limits_{i=0}^{l}\textbf{G}_{\tau k+d+i}
\right\|^2_F\right]   \label{equ:theorem1_3}\\
=& 3\eta^2 \sum\limits_{m=1}^{M}  \left[ 
\xi^{2k} d \mathbb{E}_{K(k)} \left\| \sum\limits_{i=1}^{d-1} g(x_{d-1}^{(m)}) \right\|^2
+d \mathbb{E}_{K(k)}\left\| \sum\limits_{j=1}^{k}\sum\limits_{i=0}^{\tau -1} \xi^j g(x_{\tau (k-j)+d+i}^{(m)}) \right\|^2
+ \mathbb{E}_{K(k)} \sum\limits_{l=\tau-d}^{\tau-1} \left\| \sum\limits_{i=0}^{l}g(x_{\tau k+d+i}^{(m)}) 
\right\|^2\right]  \\
=& 3\eta^2d \left[
\underbrace{ \sum\limits_{m=1}^{M}  
	\xi^{2k}  \left\| \sum\limits_{i=1}^{d-1} g(x_{d-1}^{(m)}) \right\|^2 }_{T_1}
+  \underbrace{ \sum\limits_{m=1}^{M}  
	\mathbb{E}_{K(k)}\left\| \sum\limits_{j=1}^{k} \xi^j\sum\limits_{i=0}^{\tau -1} g(x_{\tau (k-j)+d+i}^{(m)}) \right\|^2}_{T_2}
+  \underbrace{\frac{1}{d}\sum\limits_{m=1}^{M}  \sum\limits_{l=\tau-d}^{\tau-1}
	\mathbb{E}_{K(k)}\left\| \sum\limits_{i=0}^{l}g(x_{\tau k+d+i}^{(m)}) 
	\right\|^2}_{T_3}\right] 
\end{align}

where the \eqref{equ:theorem1_3} is due to the operator norm of $\textbf{J} - \textbf{I}$ is less than $1$.
For $T2$, we have
\begin{align}\nonumber
& \sum\limits_{m=1}^{M}  
\mathbb{E}_{K(k)}\left\| \sum\limits_{j=1}^{k} \xi^j\sum\limits_{i=0}^{\tau -1} g(x_{\tau (k-j)+d+i}^{(m)}) \right\|^2\\
=& \sum\limits_{m=1}^{M}  
\mathbb{E}_{K(k)}\left\| 
\sum\limits_{j=1}^{k} \xi^j\sum\limits_{i=0}^{\tau -1} \left[  g(x_{\tau (k-j)+d+i}^{(m)}) -\bigtriangledown F (x_{\tau (k-j)+d+i}^{(m)}) \right] 
+\sum\limits_{j=1}^{k} \xi^j\sum\limits_{i=0}^{\tau -1} \bigtriangledown F (x_{\tau (k-j)+d+i}^{(m)}) 
\right\|^2\\
\leq& \underbrace{2\sum\limits_{m=1}^{M}  \mathbb{E}_{K(k)}\left\| 
	\sum\limits_{j=1}^{k} \xi^j\sum\limits_{i=0}^{\tau -1} \left[  g(x_{\tau (k-j)+d+i}^{(m)}) -\bigtriangledown F (x_{\tau (k-j)+d+i}^{(m)}) \right] \right\|^2}_{T_4}
+\underbrace{2\sum\limits_{m=1}^{M}  \mathbb{E}_{K(k)}\left\| \sum\limits_{j=1}^{k} \xi^j\sum\limits_{i=0}^{\tau -1} \bigtriangledown F (x_{\tau (k-j)+d+i}^{(m)}) 
	\right\|^2 }_{T_5}
\end{align}

For the first term $T_4$, since the stochastic gradients are unbiased, all cross terms are zero. Thus, combining with Assumption 3, we have
\begin{align}\nonumber
T_4 =~& 2\sum\limits_{m=1}^{M}  \sum\limits_{j=1}^{k} \xi^{2j}\sum\limits_{i=0}^{\tau -1}\mathbb{E}_{K(k)}
\left\| g(x_{\tau (k-j)+d+i}^{(m)}) -\bigtriangledown F (x_{\tau (k-j)+d+i}^{(m)})  \right\|^2\\
\leq~& 2\sum\limits_{m=1}^{M}  \sum\limits_{j=1}^{k} \xi^{2j}\sum\limits_{i=0}^{\tau -1}
\left[ \beta \left\| \bigtriangledown F (x_{\tau (k-j)+d+i}^{(m)})  \right\|^2+\sigma^2\right] \\
=~& 2\sum\limits_{j=1}^{k} \xi^{2j}\sum\limits_{i=0}^{\tau -1}
\left[ \beta \left\| \bigtriangledown F (\textbf{X}_{\tau (k-j)+d+i})  \right\|^2_F+M\sigma^2\right] 
\end{align}

For the second term $T_5$, directly applying Jensen’s inequality, we get
\begin{align}\nonumber
T_5 =~& 2\sum\limits_{m=1}^{M}  \mathbb{E}_{K(k)}\left\| \sum\limits_{j=1}^{k} \xi^j\sum\limits_{i=0}^{\tau -1} \bigtriangledown F (x_{\tau (k-j)+d+i}^{(m)}) 
\right\|^2\\
\leq~& 2k \sum\limits_{m=1}^{M}  \mathbb{E}_{K(k)}\sum\limits_{j=1}^{k} \xi^{2j}
\left\| \sum\limits_{i=0}^{\tau -1} \bigtriangledown F (x_{\tau (k-j)+d+i}^{(m)}) 
\right\|^2\\
\leq~& 2k\tau \sum\limits_{m=1}^{M}  \mathbb{E}_{K(k)}\sum\limits_{j=1}^{k} \xi^{2j}\sum\limits_{i=0}^{\tau -1}
\left\|  \bigtriangledown F (x_{\tau (k-j)+d+i}^{(m)}) 
\right\|^2\\
=~& 2k\tau \mathbb{E}_{K(k)}\sum\limits_{j=1}^{k} \xi^{2j}\sum\limits_{i=0}^{\tau -1}
\left\|  \bigtriangledown F (\textbf{X}_{\tau (k-j)+d+i}) 
\right\|^2_F
\end{align}

Substituting the bounds of $T_4$ and $T_5$ into $T_2$
\begin{align}
T_2 \leq~& 2\sum\limits_{j=1}^{k} \xi^{2j}\sum\limits_{i=0}^{\tau -1}
\left[ \beta  \mathbb{E}\left\| \bigtriangledown F (\textbf{X}_{\tau (k-j)+d+i})  \right\|^2_F+\sigma^2\right]
+2k\tau \sum\limits_{j=1}^{k} \xi^j\sum\limits_{i=0}^{\tau -1}
\mathbb{E}\left\|  \bigtriangledown F (\textbf{X}_{\tau (k-j)+d+i}) 
\right\|^2_F\\
=~& 2\sum\limits_{j=1}^{k} \xi^{2j}M\sigma^2\tau
+(2\beta+2k\tau) \sum\limits_{j=1}^{k} \xi^j\sum\limits_{i=0}^{\tau -1}
\mathbb{E}\left\|  \bigtriangledown F (\textbf{X}_{\tau (k-j)+d+i}) 
\right\|^2_F\\
\leq~& 2M\sigma^2\tau\frac{\xi^{2}}{1-\xi^{2}}
+(2\beta+2k\tau) \sum\limits_{j=1}^{k} \xi^j\sum\limits_{i=0}^{\tau -1}
\mathbb{E}\left\|  \bigtriangledown F (\textbf{X}_{\tau (k-j)+d+i}) 
\right\|^2_F \label{equ:theorem1_4}
\end{align}

where \eqref{equ:theorem1_4} according to the summation formula of power

$$\sum\limits_{j=1}^{k} \xi^{2j}\leq\sum\limits_{j=1}^{\infty} \xi^{2j}\leq\frac{\xi^{2}}{1-\xi^{2}}$$

For $T_3$, we have
\begin{align}\nonumber
T_3 =~& \frac{1}{d}\sum\limits_{m=1}^{M}  \sum\limits_{l=\tau-d}^{\tau-1}
\mathbb{E}_{K(k)}\left\| \sum\limits_{i=0}^{l}g(x_{\tau k+d+i}^{(m)}) 
\right\|^2\\
=~& \frac{1}{d}\sum\limits_{m=1}^{M}  \sum\limits_{l=\tau-d}^{\tau-1}
\mathbb{E}_{K(k)}\left\| \sum\limits_{i=0}^{l}  \left( g(x_{\tau k+d+i}^{(m)}) -\bigtriangledown F(x_{\tau k+d+i}^{(m)})\right) 
+\sum\limits_{i=0}^{l} \bigtriangledown F(x_{\tau k+d+i}^{(m)})
\right\|^2\\
\leq~& \frac{2}{d}\sum\limits_{m=1}^{M}  \sum\limits_{l=\tau-d}^{\tau-1}
\mathbb{E}_{K(k)}\left\| \sum\limits_{i=0}^{l}  \left( g(x_{\tau k+d+i}^{(m)}) -\bigtriangledown F(x_{\tau k+d+i}^{(m)})\right) \right\|^2
+\frac{2}{d}\sum\limits_{m=1}^{M}  \sum\limits_{l=\tau-d}^{\tau-1}
\mathbb{E}_{K(k)}\left\|\sum\limits_{i=0}^{l} \bigtriangledown F(x_{\tau k+d+i}^{(m)})
\right\|^2\\
\leq~& \frac{2}{d}\sum\limits_{m=1}^{M}  \sum\limits_{l=\tau-d}^{\tau-1}\sum\limits_{i=0}^{l}
\mathbb{E}_{K(k)}\left\|  g(x_{\tau k+d+i}^{(m)}) -\bigtriangledown F(x_{\tau k+d+i}^{(m)})\right\|^2
+\frac{2}{d}\sum\limits_{m=1}^{M}  \sum\limits_{l=\tau-d}^{\tau-1}
\sum\limits_{i=0}^{l}\mathbb{E}_{K(k)}\left\|\bigtriangledown F(x_{\tau k+d+i}^{(m)})
\right\|^2\\
\leq~& \frac{2}{d}\sum\limits_{m=1}^{M}  \sum\limits_{l=\tau-d}^{\tau-1}\sum\limits_{i=0}^{l}
\left[\beta\mathbb{E}_{K(k)}\left\|  \bigtriangledown F(x_{\tau k+d+i}^{(m)}) \right\|^2 + \sigma^2\right] 
+\frac{2}{d}\sum\limits_{m=1}^{M}  \sum\limits_{l=\tau-d}^{\tau-1}\sum\limits_{i=0}^{l}
\mathbb{E}_{K(k)}\left\|\bigtriangledown F(x_{\tau k+d+i}^{(m)})
\right\|^2\\
=~& \frac{2M}{d} \sum\limits_{l=\tau-d}^{\tau-1}\sum\limits_{i=0}^{l}\sigma^2
+\frac{2\beta+2}{d}  \sum\limits_{l=\tau-d}^{\tau-1}\sum\limits_{i=0}^{l}
\mathbb{E}_{K(k)}\left\|  \bigtriangledown F(\textbf{X}_{\tau k+d+i}) \right\|^2_F 
\end{align}

We have
\begin{align}\nonumber
&\sum\limits_{i=\tau-d}^{\tau-1} \mathbb{E}_{K(k)} \left\| \textbf{X}_{\tau k+d}\textbf{J}-\textbf{X}_{\tau k+d+i}\right\|^2_F\\
=~& 3\eta^2d \left[
\sum\limits_{m=1}^{M}  
\xi^{2k}  \left\| \sum\limits_{i=1}^{d-1} g(x_{d-1}^{(m)}) \right\|^2 
+   \sum\limits_{m=1}^{M}  
\mathbb{E}_{K(k)}\left\| \sum\limits_{j=1}^{k} \xi^j\sum\limits_{i=0}^{\tau -1} g(x_{\tau (k-j)+d+i}^{(m)}) \right\|^2
+ \sum\limits_{m=1}^{M}  \sum\limits_{l=\tau-d}^{\tau-1}
\mathbb{E}_{K(k)}\left\| \sum\limits_{i=0}^{l}g(x_{\tau k+d+i}^{(m)}) 
\right\|^2\right] \\
\leq~& 3\eta^2d \left[
\sum\limits_{m=1}^{M}  
\xi^{2k}  \left\| \sum\limits_{i=1}^{d-1} g(x_{d-1}^{(m)}) \right\|^2 
+ 2M\sigma^2\tau\frac{\xi^{2}}{1-\xi^{2}}
+(2\beta+2k\tau) \sum\limits_{j=1}^{k} \xi^j\sum\limits_{i=0}^{\tau -1}
\mathbb{E}\left\|  \bigtriangledown F (\textbf{X}_{\tau (k-j)+d+i}) 
\right\|^2_F\right. \\
&+\left. \frac{2M}{d} \sum\limits_{l=\tau-d}^{\tau-1}\sum\limits_{i=0}^{l}\sigma^2
+\frac{2\beta+2}{d}  \sum\limits_{l=\tau-d}^{\tau-1}\sum\limits_{i=0}^{l}
\mathbb{E}_{K(k)}\left\|  \bigtriangledown F(\textbf{X}_{\tau k+d+i}) \right\|^2_F 
\right] 
\end{align}

And
\begin{align}\nonumber
&\sum\limits_{l=0}^{\tau-1-d} \mathbb{E}_{K(k)} \left\| \textbf{X}_{\tau k+d}\textbf{J}-\textbf{X}_{\tau k+d+l}\right\|^2_F\\
=~& 3\eta^2\left[
(\tau-d) \sum\limits_{m=1}^{M}  
\xi^{2k}  \left\| \sum\limits_{i=1}^{d-1} g(x_{d-1}^{(m)}) \right\|^2 
+  (\tau-d)  \sum\limits_{m=1}^{M}  
\mathbb{E}_{K(k)}\left\| \sum\limits_{j=1}^{k} \xi^j\sum\limits_{i=0}^{\tau -1} g(x_{\tau (k-j)+d+i}^{(m)}) \right\|^2
+ \sum\limits_{m=1}^{M}  \sum\limits_{l=0}^{\tau-1-d}
\mathbb{E}_{K(k)}\left\| \sum\limits_{i=0}^{l}g(x_{\tau k+d+i}^{(m)}) 
\right\|^2\right] \\
\leq~& 3\eta^2(\tau-d)  \left[
\sum\limits_{m=1}^{M}  
\xi^{2k}  \left\| \sum\limits_{i=1}^{d-1} g(x_{d-1}^{(m)}) \right\|^2 
+ 2M\sigma^2\tau\frac{\xi^{2}}{1-\xi^{2}}
+(2\beta+2k\tau) \sum\limits_{j=1}^{k} \xi^j\sum\limits_{i=0}^{\tau -1}
\mathbb{E}\left\|  \bigtriangledown F (\textbf{X}_{\tau (k-j)+d+i}) 
\right\|^2_F\right. \\
&+\left. \frac{2M}{\tau-d} \sum\limits_{l=0}^{\tau-1-d}\sum\limits_{i=0}^{l}\sigma^2
+\frac{2\beta+2}{\tau-d}  \sum\limits_{l=0}^{\tau-1-d}\sum\limits_{i=0}^{l}
\mathbb{E}_{K(k)}\left\|  \bigtriangledown F(\textbf{X}_{\tau k+d+i}) \right\|^2_F 
\right] 
\end{align}

Then, summing over all periods from $k= 0$ to $k= K$, where $K$ is the total global iterations:
\begin{align}
&\sum\limits_{k=1}^{K} \sum\limits_{i=\tau-d}^{\tau-1} \mathbb{E}_{K(k)} \left\| \textbf{X}_{\tau k+d}\textbf{J}-\textbf{X}_{\tau k+d+i}\right\|^2_F\\
\leq~& 3\eta^2d \sum\limits_{k=1}^{K}\left[
\sum\limits_{m=1}^{M}  
\xi^{2k}  \left\| \sum\limits_{i=1}^{d-1} g(x_{d-1}^{(m)}) \right\|^2 
+ 2M\sigma^2\tau\frac{\xi^{2}}{1-\xi^{2}}
+(2\beta+2k\tau) \sum\limits_{j=1}^{k} \xi^j\sum\limits_{i=0}^{\tau -1}
\mathbb{E}\left\|  \bigtriangledown F (\textbf{X}_{\tau (k-j)+d+i}) 
\right\|^2_F\right. \\
&+\left. \frac{2M}{d} \sum\limits_{l=\tau-d}^{\tau-1}\sum\limits_{i=0}^{l}\sigma^2
+\frac{2\beta+2}{d}  \sum\limits_{l=\tau-d}^{\tau-1}\sum\limits_{i=0}^{l}
\mathbb{E}_{K(k)}\left\|  \bigtriangledown F(\textbf{X}_{\tau k+d+i}) \right\|^2_F 
\right] \\
\leq &3\eta^2d \frac{\xi^{2}}{1-\xi^{2}}   \left\| \sum\limits_{i=1}^{d-1} g(\textbf{X}_{d-1}) \right\|^2_F 
+ 6\eta^2dKM\sigma^2\tau\frac{\xi^{2}}{1-\xi^{2}}
+ \frac{6\eta^2d MK}{d} \sum\limits_{l=\tau-d}^{\tau-1}\sum\limits_{i=0}^{l}\sigma^2\\
&+3\eta^2d(2\beta+2k\tau)\sum\limits_{k=1}^{K} \sum\limits_{j=1}^{k} \xi^j\sum\limits_{i=0}^{\tau -1}
\mathbb{E}\left\|  \bigtriangledown F (\textbf{X}_{\tau (k-j)+d+i}) 
\right\|^2_F 
+ 3\eta^2(2\beta+2) \sum\limits_{k=1}^{K}\sum\limits_{l=\tau-d}^{\tau-1}\sum\limits_{i=0}^{l}
\mathbb{E}\left\|  \bigtriangledown F(\textbf{X}_{\tau k+d+i}) \right\|^2_F \label{equ:theorem1_7}
\end{align}

Expanding the summation, we have
\begin{align}
\sum\limits_{k=1}^{K} \sum\limits_{j=1}^{k} \xi^j\sum\limits_{i=0}^{\tau -1}
\mathbb{E}\left\|  \bigtriangledown F (\textbf{X}_{\tau (k-j)+d+i}) 
\right\|^2_F 
=~&\sum\limits_{k=1}^{K}  \sum\limits_{r=0}^{k-1}  \left[\xi^{k-r}\sum\limits_{i=0}^{\tau -1}
\mathbb{E}\left\|  \bigtriangledown F (\textbf{X}_{\tau r+d+i}) 
\right\|^2_F\right] \\
\leq~&\sum\limits_{r=1}^{K}  \left[ \left( \sum\limits_{i=0}^{\tau -1}
\mathbb{E}\left\|  \bigtriangledown F (\textbf{X}_{\tau r+d+i}) 
\right\|^2_F\right) 
\left( \sum\limits_{k=r}^{K}\xi^{k-r}\right) 
\right]  \\
\leq~&\sum\limits_{r=1}^{K}  \left[ \left( \sum\limits_{i=0}^{\tau -1}
\mathbb{E}\left\|  \bigtriangledown F (\textbf{X}_{\tau r+d+i}) 
\right\|^2_F\right) 
\left( \sum\limits_{k=r}^{+\infty}\xi^{k-r}\right) 
\right]  \\
\leq~&\frac{1}{1-\xi}\sum\limits_{k=1}^{K} \sum\limits_{i=0}^{\tau -1}
\mathbb{E}\left\|  \bigtriangledown F (\textbf{X}_{\tau k+d+i}) 
\right\|^2_F \label{equ:theorem1_5}
\end{align}

Thus, we have
\begin{align}
\sum\limits_{k=1}^{K}\sum\limits_{l=\tau-d}^{\tau-1}\sum\limits_{i=0}^{l}
\mathbb{E}\left\|  \bigtriangledown F(\textbf{X}_{\tau k+d+i}) \right\|^2_F 
\leq d\sum\limits_{k=1}^{K}
\sum\limits_{i=0}^{\tau -1}
\mathbb{E}\left\|  \bigtriangledown F(\textbf{X}_{\tau k+d+i}) \right\|^2_F\label{equ:theorem1_6}
\end{align}

Plugging \eqref{equ:theorem1_5} and \eqref{equ:theorem1_6} into \eqref{equ:theorem1_7},
\begin{align}\nonumber
&\sum\limits_{k=1}^{K} \sum\limits_{i=\tau-d}^{\tau-1} \mathbb{E}_{K(k)} \left\| \textbf{X}_{\tau k+d}\textbf{J}-\textbf{X}_{\tau k+d+i}\right\|^2_F\\
\leq &3\eta^2d \frac{\xi^{2}}{1-\xi^{2}}   \left\| \sum\limits_{i=1}^{d-1} g(\textbf{X}_{d-1}) \right\|^2_F 
+ 6\eta^2dKM\sigma^2\tau\frac{\xi^{2}}{1-\xi^{2}}
+ \frac{6\eta^2d MK}{d} \sum\limits_{l=\tau-d}^{\tau-1}\sum\limits_{i=0}^{l}\sigma^2\\
&+3\eta^2d\frac{2\beta+2k\tau}{1-\xi}\sum\limits_{k=1}^{K} \sum\limits_{i=0}^{\tau -1}
\mathbb{E}\left\|  \bigtriangledown F (\textbf{X}_{\tau k+d+i}) 
\right\|^2_F
+ 3\eta^2(2\beta+2) d\sum\limits_{k=1}^{K} \sum\limits_{i=0}^{\tau -1}
\mathbb{E}\left\|  \bigtriangledown F(\textbf{X}_{\tau k+d+i}) \right\|^2_F
\end{align}

And
\begin{align}\nonumber
&\sum\limits_{k=1}^{K}\sum\limits_{l=0}^{\tau-1-d} \mathbb{E}_{K(k)} \left\| \textbf{X}_{\tau k+d}\textbf{J}-\textbf{X}_{\tau k+d+l}\right\|^2_F\\
\leq~& 3\eta^2(\tau-d) \frac{\xi^{2}}{1-\xi^{2}}   \left\| \sum\limits_{i=1}^{d-1} g(\textbf{X}_{d-1}) \right\|^2_F 
+ 6\eta^2(\tau-d) KM\sigma^2\tau\frac{\xi^{2}}{1-\xi^{2}}
+6\eta^2MK\sum\limits_{l=0}^{\tau-1-d}\sum\limits_{i=0}^{l}\sigma^2\\
&
+3\eta^2(\tau-d) \frac{2\beta+2k\tau}{1-\xi} \sum\limits_{k=1}^{K} \sum\limits_{i=0}^{\tau -1}
\mathbb{E}\left\|  \bigtriangledown F (\textbf{X}_{\tau k+d+i}) 
\right\|^2_F
+3\eta^2(2\beta+2)(\tau-d) \sum\limits_{k=1}^{K}
\sum\limits_{i=0}^{\tau-1-d}
\mathbb{E}\left\|  \bigtriangledown F(\textbf{X}_{\tau k+d+i}) \right\|^2_F 
\end{align}

Recall the intermediate result \eqref{equ:lemma3_4} in the proof of Lemma 1:
\begin{align}
&\mathbb{E}_{K(k)}\left[ \frac{1}{K}\sum\limits_{k=1}^{K}  
\left\|  \bigtriangledown F(\mu_{K(k)})\right\|^2 \right] \\
\leq &\frac{2\left[ F(\mu_1) - F_{inf}\right] }{\eta K(\xi d+\tau-d)} 
+ \frac{2L\eta\sigma^2 \left[ \xi^2 d + \tau-d \right]}{M(\xi d+\tau-d)} \\
& + \frac{\eta^2 L^2 }{KM(\xi d+\tau-d ) } \left[  3\eta^2(\tau-d) \frac{\xi^{2}}{1-\xi^{2}}   \left\| \sum\limits_{i=1}^{d-1} g(\textbf{X}_{d-1}) \right\|^2_F 
+ 6\eta^2 MK   \left( (\tau-d)\sigma^2\tau\frac{\xi^{2}}{1-\xi^{2}}+\sum\limits_{l=0}^{\tau-1-d}\sum\limits_{i=0}^{l}\sigma^2    \right) \right] \\
&+\frac{\eta^2 \xi L^2}{KM(\xi d+\tau-d )} \left[   3\eta^2d \frac{\xi^{2}}{1-\xi^{2}}   \left\| \sum\limits_{i=1}^{d-1} g(\textbf{X}_{d-1}) \right\|^2_F 
+ 6\eta^2MK  \left(  d\sigma^2\tau\frac{\xi^{2}}{1-\xi^{2}}
+ \sum\limits_{l=\tau-d}^{\tau-1}\sum\limits_{i=0}^{l}\sigma^2 \right)  \right] \\
&+ \frac{2L\xi^2\eta^4(\beta+1)-\eta^2\xi M}{KM^2 (\xi d+\tau-d )}  
\sum\limits_{k=1}^{K} \sum\limits_{i=\tau-d}^{\tau-1}  \mathbb{E}_{K(k)}\left\| \bigtriangledown F\left(  \textbf{X}_{\tau k+d+i}\right) \right\|^2_F \label{equ:theorem1_11} \\
&+ \frac{2L\xi^2\eta^4(\beta+1)-\eta^2 M}{KM^2 (\xi d+\tau-d )} 
\sum\limits_{k=1}^{K} \sum\limits_{i=0}^{\tau-1-d}  \mathbb{E}_{K(k)} \left\| \bigtriangledown F\left(  \textbf{X}_{\tau k+d+i}\right) \right\|^2_F \label{equ:theorem1_10}\\
&  + \frac{\eta^2 L^2 }{KM(\xi d+\tau-d ) } \left[ 
3\eta^2(\tau-d) \frac{2\beta+2k\tau}{1-\xi} \sum\limits_{k=1}^{K} \sum\limits_{i=0}^{\tau -1}
\mathbb{E}\left\|  \bigtriangledown F (\textbf{X}_{\tau k+d+i}) 
\right\|^2_F
+3\eta^2(2\beta+2)(\tau-d) \sum\limits_{k=1}^{K}
\sum\limits_{i=0}^{\tau-1-d}
\mathbb{E}\left\|  \bigtriangledown F(\textbf{X}_{\tau k+d+i}) \right\|^2_F \right]\label{equ:theorem1_9} \\
&+\frac{\eta^2 \xi L^2}{KM(\xi d+\tau-d )} \left[ 
3\eta^2d\frac{2\beta+2k\tau}{1-\xi}\sum\limits_{k=1}^{K} \sum\limits_{i=0}^{\tau -1}
\mathbb{E}\left\|  \bigtriangledown F (\textbf{X}_{\tau k+d+i}) 
\right\|^2_F
+ 3\eta^2(2\beta+2) d\sum\limits_{k=1}^{K} \sum\limits_{i=0}^{\tau -1}
\mathbb{E}\left\|  \bigtriangledown F(\textbf{X}_{\tau k+d+i}) \right\|^2_F \right] \label{equ:theorem1_8}
\end{align}

Rearrange \eqref{equ:theorem1_11}, \eqref{equ:theorem1_10}, \eqref{equ:theorem1_9}, and \eqref{equ:theorem1_8}
\begin{equation}\begin{aligned}\nonumber
& \frac{2L\xi^2\eta^4(\beta+1)/M-\eta^2\xi }{KM (\xi d+\tau-d )}  
\sum\limits_{k=1}^{K} \sum\limits_{i=\tau-d}^{\tau-1}  \mathbb{E}_{K(k)}\left\| \bigtriangledown F\left(  \textbf{X}_{\tau k+d+i}\right) \right\|^2_F  \\
&+ \left( \frac{ 2L\eta\xi^2\eta^3(\beta+1)(1-\xi)-\eta^2+ 3L^2\eta^4(2\beta+2)(\tau-d)(1-\xi)             }{KM(\xi d+\tau-d )(1-\xi) } 
\right) 
\sum\limits_{k=1}^{K} \sum\limits_{i=0}^{\tau-1-d}  \mathbb{E}_{K(k)} \left\| \bigtriangledown F\left(  \textbf{X}_{\tau k+d+i}\right) \right\|^2_F \\
&  + \left( \frac{\eta^2 L^2 3\eta^2(\tau-d) (2\beta+2k\tau)    +  3d\eta^4 \xi L^2 [ (2\beta+2k\tau)+(2\beta+2)(1-\xi) ]    }{KM(\xi d+\tau-d ) (1-\xi)} 
\right) \sum\limits_{k=1}^{K} \sum\limits_{i=0}^{\tau -1}
\mathbb{E}\left\|  \bigtriangledown F (\textbf{X}_{\tau k+d+i}) 
\right\|^2_F
\end{aligned}\end{equation}

When the learning rate satisfies the following two formulas at the same time
\begin{align}\nonumber
\eta  \leq & \sqrt{ \frac{1}{2L\xi^2(\beta+1)(1-\xi)+ 3L^2(\tau-d)[(1-\xi)(2\beta+2) +(2\beta+2k\tau)    ]+  3d \xi L^2 [ (2\beta+2k\tau)+(2\beta+2)(1-\xi) ] }}\\
= & \sqrt{ \frac{1}{2L\xi^2(\beta+1)(1-\xi)+  6 L^2 (d \xi+\tau -d)[ (\beta+k\tau)+(\beta+1)(1-\xi) ] }}
\end{align}

\begin{align}\nonumber
\eta  \leq & \sqrt{ \frac{\xi M(1-\xi)}{2L\xi^2(\beta+1)(1-\xi)+ 3L^2 M(\tau-d) (2\beta+2k\tau)    +  6dM \xi L^2 [ (\beta+k\tau)+(\beta+1)(1-\xi) ]  }}
\end{align}

We have
\begin{align}\nonumber
&\mathbb{E}_{K(k)}\left[ \frac{1}{K}\sum\limits_{k=1}^{K}  
\left\|  \bigtriangledown F(\mu_{K(k)})\right\|^2 \right] \\
\leq &\frac{2\left[ F(\mu_1) - F_{inf}\right] }{\eta K(\xi d+\tau-d)} 
+ \frac{2L\eta\sigma^2 \left[ \xi^2 d + \tau-d \right]}{M(\xi d+\tau-d)} \\
& + \frac{\eta^4 L^2 }{KM(\xi d+\tau-d ) } \left[  3(\tau-d) \frac{\xi^{2}}{1-\xi^{2}}   \left\| \sum\limits_{i=1}^{d-1} g(\textbf{X}_{d-1}) \right\|^2_F 
+ 6MK   \left( (\tau-d)\sigma^2\tau\frac{\xi^{2}}{1-\xi^{2}}+\sum\limits_{l=0}^{\tau-1-d}\sum\limits_{i=0}^{l}\sigma^2    \right) \right] \\
&+\frac{\eta^4 \xi L^2}{KM(\xi d+\tau-d )} \left[   3d \frac{\xi^{2}}{1-\xi^{2}}   \left\| \sum\limits_{i=1}^{d-1} g(\textbf{X}_{d-1}) \right\|^2_F 
+ 6MK  \left(  d\sigma^2\tau\frac{\xi^{2}}{1-\xi^{2}}
+ \sum\limits_{l=\tau-d}^{\tau-1}\sum\limits_{i=0}^{l}\sigma^2 \right)  \right] \\
\leq &\frac{2M\left[ F(\mu_1) - F_{inf}\right]   
	+   2MKL\eta^2\sigma^2 \left[ \xi^2 d + \tau-d \right]   }{\eta MK(\xi d+\tau-d)}  +\frac{3\eta^4 \xi L^2  (\tau-d+d\xi )   }{MK(\xi d+\tau-d )} \frac{\xi^{2}}{1-\xi^{2}}   \left\| \sum\limits_{i=1}^{d-1} g(\textbf{X}_{d-1}) \right\|^2_F  \\
&+ \frac{6 \eta^4 L^2 \sigma^2  }{\xi d+\tau-d  }   \left( 
\tau\frac{\xi^{2}}{1-\xi^{2}}(\tau-d+\xi d)
+(\tau-d)^2+\xi d(\tau -1)
\right) 
\end{align}

\bigskip \hrule  \bigskip

\textbf{Corollary 1.} 
Under sssumptions, if the learning rate is $\eta= A/\sqrt{K}$ the average-squared gradient norm after $K$ iterations is bounded by
\begin{equation}\begin{aligned}\nonumber
&\mathbb{E}_{K(k)}\left[ \frac{1}{K}\sum\limits_{k=1}^{K}  
\left\|  \bigtriangledown F(\mu_{K(k)})\right\|^2 \right] \\
\leq &\frac{2M\left[ F(\mu_1) - F_{inf}\right]   
	+   2MLA^2\sigma^2 \left[ \xi^2 d + \tau-d \right]   }{AM\sqrt{K}(\xi d+\tau-d)}  +\frac{3A^4 \xi L^2  (\tau-d+d\xi )   }{MK^3(\xi d+\tau-d )} \frac{\xi^{2}}{1-\xi^{2}}   \left\| \sum\limits_{i=1}^{d-1} g(\textbf{X}_{d-1}) \right\|^2_F  \\
&+ \frac{6 A^4 L^2 \sigma^2  }{K^2(\xi d+\tau-d)  }   \left( 
\tau\frac{\xi^{2}}{1-\xi^{2}}(\tau-d+\xi d)
+(\tau-d)^2+\xi d(\tau -1)
\right) 
\end{aligned}\end{equation}

If the total iterations $K$ is sufficiently large, then the average-squared gradient norm will be bounded by 
$$\mathbb{E}_{K(k)}\left[ \frac{1}{K}\sum\limits_{k=1}^{K}  
\left\|  \bigtriangledown F(\mu_{K(k)})\right\|^2 \right] 
\leq
\frac{2M\left[ F(\mu_1) - F_{inf}\right]   
	+   2MLA^2\sigma^2 \left[ \xi^2 d + \tau-d \right]   }{AM\sqrt{K}(\xi d+\tau-d)} $$.